\documentclass[twocolumn,aps,showpacs,pra]{revtex4-1}
\usepackage{amsmath}
\usepackage{amssymb}
\usepackage{esint}
\usepackage{graphicx}
\usepackage{bbold}
\usepackage[normalem]{ulem}
\usepackage[colorlinks=true,allcolors=blue,bookmarks=false,pdfusetitle]{hyperref}
\usepackage{url}

\makeatletter

\usepackage{color}

\newcommand{\sgn}{\mbox{sgn}}
\newcommand{\dert}{\partial_t}
\newcommand{\derx}{\partial_x}

\makeatother
\begin{document}
\title{Finite-temperature dynamics of a Tonks-Girardeau gas in a frequency-modulated harmonic trap}
\author{Y. Y. Atas}
\address{School of Mathematics and Physics, University of Queensland, Brisbane, Queensland 4072, Australia}
\author{S. A. Simmons}
\address{School of Mathematics and Physics, University of Queensland, Brisbane, Queensland 4072, Australia}
\author{K. V. Kheruntsyan}
\address{School of Mathematics and Physics, University of Queensland, Brisbane, Queensland 4072, Australia}

\date{\today}
\begin{abstract}
We study the out-of-equilibrium dynamics of a finite-temperature harmonically trapped Tonks-Girardeau gas induced by periodic modulation of the trap frequency. We give explicit exact solutions for the real-space density and momentum distributions of this interacting many-body system and characterise the stability diagram of the dynamics by mapping the many-body solution to the solution and stability diagram of Mathieu's differential equation. The mapping allows one to deduce the exact structure of parametric resonances in the parameter space characterized by the driving amplitude and frequency of the modulation. Furthermore, we analyse the same problem within the finite-temperature hydrodynamic approach and show that the respective solutions to the hydrodynamic equations can be mapped to the same Mathieu equation. Accordingly, the stability diagram and the structure of resonances following from the hydrodynamic approach is exactly the same as those obtained from the exact many-body solution.
\end{abstract}
\maketitle

\section{Introduction}
Characterising and understanding the behaviour of quantum many-body systems driven out-of-equilibrium is one of the grand challenges of modern physics. The problem is particularly challenging when treating the dynamics of realistic finite-temperature systems, rather than systems evolving from the zero-temperature ground state. Here we solve this problem for a paradigmatic model in many-body physics---a Tonks-Girardeau (TG) gas \cite{Girardeau1960} of bosons interacting in one dimension (1D) via hard-core repulsion. The specific dynamical protocol that we consider is periodic modulation of the frequency of the harmonic trapping potential, which continuously drives the system out-of-equilibrium. Periodically driven systems are rather common in nature and form an important class of problems in quantum dynamics. In the present context, a periodically driven TG gas can be realised in ultra-cold atom experiments \cite{paredes2004tonks,Kinoshita2004,Kinoshita2006,Naagerl2009,Naagerl2015}, and therefore our results are of direct relevance to the ongoing experimental progress in ultra-cold atom physics, aimed at developing a fundamental understanding of out-of-equilibrium phenomena in many-body physics \cite{Bloch-Dalibard-Zwerger,CazalillaRigolNJP2010,polkovnikov2011,Cazalilla2011,lamacraft2012}.

Previously, a harmonically trapped single particle has been widely studied and the effect of modulating either the position of the trap center or the strength of the trapping potential has been thoroughly investigated \cite{Husimi1953,Breuer1989,He-Brown2014,Brown1991,Hanggi1993,Kohler-Hanggi1997,Zerbe-Hanggi1994,Thorwart-Hanggi2000}.
More recently, in the context of many-body out-of-equilibrium dynamics, the periodic modulation of the trapping frequency of a one-dimensional system of interacting bosons at zero temperature was considered by Quinn and Haque \cite{QuinnHaque2014}. In particular, they focussed primarily on the energy absorption of the system and the structure of energy resonances for interaction strengths varying from no interactions through to the strongly interacting TG regime. 

In this work, we extend the analysis of Quinn and Haque in the TG regime to characterise the non-equilibrium dynamics of the gas at finite temperatures and show how the solution to this many-body problem can be mapped to the solution of Mathieu's differential equation \cite{Mathieu1868,mclachlan1964theory}.
The approach we use has been developed recently in Ref.~\cite{Atas2017a} and provides an exact finite-temperature theory that can in general model the dynamics of a TG gas in arbitrary trapping potentials. So far this treatment has been applied to a harmonically trapped TG gas driven out of equilibrium via a quench of the trap frequency from an initial value to a fixed final value \cite{Atas2017b}. Such a quench protocol invokes familiar breathing-mode oscillations of the density profile, however, an unexpected many-body effect---dubbed many-body bounce---was also found in the dynamics of the momentum distribution. The many-body bounce manifests itself as an additional narrowing of the momentum distribution during the breathing mode oscillation cycle and is similar to the phenomenon of frequency doubling first observed in a weakly interacting 1D Bose gas \cite{Fang2014,Bouchoule2016}.

The mapping to Mathieu's equation is facilitated through a scaling solution that exist for the nonequilibrium dynamics of the TG gas in a time-varying harmonic trap. \cite{GangardtMinguzziExact,QuinnHaque2014,Atas2017a,Atas2017b}. Such a scaling solution allows the evolution of the density and momentum distributions of the gas to be determined through a single scaling parameter, which itself satisfies an ordinary second-order nonlinear differential equation, the Ermakov-Pinney equation \cite{Ermakov,Pinney}. For a specific case of sinusoidal modulation of the trap frequency, the solutions to the Ermakov-Pinney equation can be found through the mapping to Mathieu's equation. Mathieu's equation itself possesses stable and unstable solutions, with a nontrivial structure of parametric resonances, which ultimately determine the dynamics and the stability diagram of the many-body TG gas that we analyse here in detail.

In addition to the exact many-body analysis, we study the dynamics of the TG gas using a finite-temperature hydrodynamic approach developed in Ref.~\cite{Bouchoule2016}. Solutions to the hydrodynamic equations are facilitated through the same scaling parameter (satisfying the same 
Ermakov-Pinney equation) as that in the exact many-body theory \cite{Ermakov,Pinney}. Accordingly, the mapping to Mathieu's equation, the stability diagram and the structure of parametric resonances that follow from the hydrodynamic approach are also the same. Thus, the frequency-modulation of a harmonically trapped Tonks-Girardeau gas represents yet another dynamical scenario, in addition to a simple quench protocol studied in Ref.~\cite{Atas2017b}, which can be adequately described by such a theory.


\section{The model and scaling solutions} \label{The_model_and_scaling_solution}

\subsection{Exact many-body solution in a time-dependent trapping potential}

The model system we consider is a 1D gas of $N$ impenetrable (hard-core) bosons \cite{Girardeau1960,Lieb-Liniger1963} of mass $m$ confined by a one-body time-dependent trapping potential $V(x,t)$. (To keep the discussion general, we do not specify the form of $V(x,t)$ until the specifics of harmonic trapping come into play.) Due to the Fermi-Bose gas mapping \cite{Girardeau1960,GirardeauWright2000,Das-Girardeau2002}, to every $N$-particle wavefunction of the bosonic problem, $\Psi^B_{N,k}(x_1,...,x_N;t)$ (where the index $k$ enumerates the different states that may occur in the given $N$-particle sector), there corresponds an antisymmetric wavefunction $\Psi^F_{N,k}(x_1,...,x_N;t)$ for a similarly trapped system of noninteracting spinless fermions, and vice versa:
\begin{equation}
\Psi^B_{N,k}(x_{1},...,x_{N};t)\!=\!A(x_{1},...,x_{N})
\Psi^F_{N,k}(x_{1},...,x_{N};t), 
\label{Bose_Fermi_mapping}
\end{equation} 
where the unit antisymmetric function
$A(x_{1},...,x_{N})\!=\!\prod_{1\leq j<i \leq N}\sgn(x_{i}-x_{j})$ ensures the symmetrization of the bosonic wavefunction. Accordingly, observables that do not depend on the sign of the many-body wave functions are readily given by their fermionic counterparts \cite{Das-Girardeau2002}, whereas those that depend on the sign behave differently.

The free-fermion $N$-particle wavefunction itself is constructed as
a Slater determinant
$\Psi_{N,k}^{F}(x_{1},...,x_{N};t)\!=\!\mathrm{det}_{i,j=1}^{N}\left[
  \phi_{k_i}(x_{j},t)\right]/{\sqrt{N!}}$
of single-particle wavefunctions $\phi_{k_i} (x,t)$ evolving according to
the time-dependent Schr\"{o}dinger equation, 
\begin{equation}
i\hbar \frac{\partial  \phi_{k_i}(x,t)}{\partial t}=-\frac{\hbar^{2}}{2m}\frac{\partial^{2} \phi_{k_i}(x,t)}{\partial x^2}+V(x,t)\phi_{k_i}(x,t),
\label{eq:Schroedinger}
\end{equation} 
with the initial wavefunctions
$\phi_{k_i} (x,0)$ being the eigenstates of the trapping potential
$V(x,0)$, with eigenenergies $E_{k_i}$ such that the total energy
$E_k\!=\!\sum_{i=1}^{N}E_{k_i}$. Thus, the index
$k\!=\!\{k_1,...,k_N\}$ in $\Psi^{B/F}_{N,k}$ labels the different sets of single-particle
quantum numbers $k_i$ that may occur in the initial $N$-particle state. The initial $N$-particle state is described by $\Psi_{N,k}^{F}(x_{1},...,x_{N};0)$, which itself evolves according to the $N$-particle Schr\"{o}dinger equation, 
\begin{equation}
i\hbar \frac{\partial  \Psi^F_{N,k}}{\partial t}=\hat{H}\,\Psi^F_{N,k},
\label{eq:Schroedinger-N}
\end{equation} 
where
\begin{equation}
\hat{H}=\sum_{i=1}^{N} \left[-\frac{\hbar^{2}}{2m} \frac{\partial^{2}}{\partial x_{i}^2}+V(x_{i},t)\right].
\label{eq:Hamiltonian-free}
\end{equation}

We next assume preparation of the system such that its
initial state at time $t\!=\!0$, in the trapping potential $V(x,0)$, is described by a grand-canonical ensemble at temperature $T_0$ and chemical potential $\mu_0$. This defines the statistical weights $P_{N,k}\!=\!\frac{1}{\mathcal{Z} }e^{(\mu_0 N- E_{k} )/k_BT_0}\geq 0$ attached to the occurrence of states $\Psi^B_{N,k}(x_1,...,x_N;0)$ in the ensemble, where  $\mathcal{Z}=\!\sum_{N,k} e^{(\mu_0 N- E_{k} )/k_BT_0}$ is the grand-canonical partition function. 
With the initial state preparation specified, the Fermi-Bose mapping allows one to reduce the problem of treating finite-temperature dynamics of a trapped TG gas to a single-particle basis. In particular, combining such mapping with the Fredholm determinant approach of Ref.~\cite{Atas2017a} leads to a computationally efficient way of calculating exactly the reduced one-body density matrix of the system, $\rho(x,y,t)$, and hence important physical observables of the gas, such as its real-space density 
\begin{equation}
\rho(x,t) \!=\! \rho(x,x;t),
\end{equation}
 and the momentum distribution
 \begin{equation}
 n(k,t)\!=\!\iint dx\, dy \, e^{-ik(x-y)}\rho(x,y;t).
 \end{equation}

 More specifically, the reduced one-body density matrix can be computed as a simple double sum \cite{Atas2017a}, 
 \begin{equation}
\rho(x,y;t)=\sum_{i,j=0}^{\infty}\sqrt{f_{i}}
\phi_{i}(x,t)Q_{ij}(x,y;t)\sqrt{f_{j}}\phi_{j}^{\ast}(y,t), 
\label{Finite_temp_densitymat_Tonks}
\end{equation}
where $Q_{ij}$ are the  
matrix elements of the operator 
$\mathbf{Q}(x,y;t)=(\mathbf{P}^{-1})^{\mathsf{T}}\mathrm{det}\;\!\mathbf{P}$, with
\begin{equation}
 P_{ij}(x,y;t)\! =\! \delta_{ij}- 2\;\!\sgn(y-x)\sqrt{f_{i}f_{j}} \!\int_{x}^{y}
 \!\!\!dx^{\prime}\phi_{i}(x^{\prime}\!,t)\phi_{j}^{\ast}(x^\prime\!,t),
  \label{matrix_element}
\end{equation}
and $f_i$ are the Fermi-Dirac occupancies $f_{i}\!=\![e^{(E_{i}-\mu)/k_BT_0}+1]^{-1}$ of single-particle orbitals $\phi_i(x,0)$.

\subsection{Scaling solution in a harmonic trap} \label{Scaling_solution_in_a_harmonic_trap}

In general, the calculation of the one-body density
matrix using Eq.~(\ref{Finite_temp_densitymat_Tonks}) requires the evaluation of the overlap matrix elements $P_{ij}(x,y;t)$
between the time-evolved single-particle wavefunctions $\phi_j(x,t)$, starting from the initial wavefunctions
$\phi_j(x,0)$. For the case of a harmonic trap, 
\begin{equation}
V(x,t)=m\omega^{2}(t)x^2/2, \label{harmonic_trapping_potential}
\end{equation}
the initial wavefunctions
$\phi_j(x,0)$ are given by the well-known Hermite-Gauss orbitals for the initial trap of frequency $\omega_0\!=\!\omega(0)$, with the 1D harmonic oscillator energy eigenvalues given by $E_{j}=\hbar \omega_{0}(j+\frac{1}{2})$, whereas
the evolved wavefunctions, for arbitrary $\omega(t)$, can be found exactly using a scaling transformation \cite{popov1969parametric,PerelomovBook,GangardtMinguzziExact},
\begin{equation}
\phi_{j}(x,t)=\frac{1}{\sqrt{\lambda}}\phi_{j}\left(\frac{x}{\lambda},0\right)\exp\left[ i \frac{mx^{2}}{2\hbar}\frac{\dot{\lambda}}{\lambda}-iE_{j}(t)t\right],
\label{eq:scaling}
\end{equation}
where $E_j(t)\!=\!\hbar \omega_{0}(j+\frac{1}{2}) \frac{1}{t}\int_{0}^{t}\mathrm{d}t^{\prime}/\lambda^{2}(t^{\prime})$, and the scaling function $\lambda(t)$ satisfies a second-order nonlinear differential equation known as the Ermakov-Pinney equation \cite{Ermakov,Pinney},
\begin{equation}
\ddot{\lambda}+\omega^{2}(t)\lambda=\frac{\omega^{2}(0)}{\lambda^{3}},
\label{ermakov}
\end{equation}
with the initial conditions $\lambda(0)\!=\!1$ and $\dot{\lambda}(0)\!=\!0$.

The scaling transformation (\ref{eq:scaling}) leads to a simple scaling solution for the evolution of the reduced one-body density matrix \cite{Atas2017a,Atas2017b},
\begin{equation}
\rho(x,y;t)=\frac{1}{\lambda}
\rho_0\left(x/\lambda,y/\lambda\right) e^{i m\dot\lambda (x^2-y^2)/2\hbar \lambda}
\label{eq.ss},
\end{equation}
where $\rho_0(x,y)\!=\!\rho(x,y;0)$ is the one-body density matrix for the initial thermal equilibrium state. Such a scaling solution simplifies the analysis enormously as the one-body density matrix needs to be calculated only once, at time $t\!=\!0$, while the specifics of the dynamics is reduced to the solution of the Ermakov-Pinney equation for a single scaling parameter $\lambda(t)$

As has been shown by Pinney \cite{Pinney}, the general solution to the Ermakov-Pinney equation (\ref{ermakov}) can be constructed by combining two independent solutions $\big{(}\lambda_{1}(t),\lambda_{2}(t)\big{)}$ of the respective homogeneous equation
\begin{equation}
\ddot{\lambda}+\omega^{2}(t)\lambda=0, 
\label{homoegeneous_ermakov}
\end{equation}
which describes a simple harmonic oscillator with time-dependent frequency. More specifically, the general solution is constructed as 
\begin{equation}
\lambda(t)=\sqrt{\mathcal{A}\lambda_{1}^{2}(t)+\mathcal{B}\lambda_{2}^{2}(t)+2\mathcal{C}\lambda_{1}(t)\lambda_{2}(t)} 
\label{general_solution_ermakov}
\end{equation}
where the constants $\mathcal{A},~\mathcal{B}$, and $\mathcal{C}$ satisfy the constraint equation $\mathcal{A}\mathcal{B}-\mathcal{C}^{2}=\omega^{2}(0)/W^{2}$, and where $W=\lambda_{1}\dot{\lambda}_{2}-\lambda_{2}\dot{\lambda}_{1}$ is the Wronskian (which is a constant, in accordance with Abel's identity) of the two independent solutions. This constraint equation, together with the initial conditions on the scaling function, $\lambda(0)\!=\!1$ and $\dot{\lambda}(0)\!=\!0$, uniquely fix the coefficients $\mathcal{A}$, $\mathcal{B}$ and $\mathcal{C}$. Note that the function $\lambda$ must be real and positive in order to keep the probability densities $|\phi_{j}(x,t)|^{2}$ positive, and as such the positive branch of the square root is assumed in (\ref{general_solution_ermakov}).

In Sec. \ref{Mapping_to_Mathieu_eq} below we will consider a specific dynamical protocol in which the TG gas evolves under sinusoidal modulation of the harmonic trap frequency $\omega(t)$. In this case, the homogeneous differential equation (\ref{homoegeneous_ermakov}) can be mapped to Mathieu's equation. Solutions to Mathieu's equation and their stability properties are known and will be presented in Sec.~\ref{Solutions to Mathieu's equation}, hence allowing us to construct the solutions to the Ermakov-Pinney equation and ultimately analyse the dynamics of the density and momentum distributions of the TG gas. Before doing so, however, we momentarily pause to point out that evolution of a harmonically trapped TG gas under the same dynamical protocol can be also analysed using a finite-temperature hydrodynamic approach developed in Ref.~\cite{Bouchoule2016,Atas2017b}.

\subsection{Hydrodynamic treatment}

 In the hydrodynamic approach, the TG gas evolves according to the following equations for the local 1D density, $\rho(x,t)$, the hydrodynamic velocity, $v(x,t)$, and the entropy per particle, $s(x,t)$~\cite{Bouchoule2016,Atas2017b}:
\begin{align}
	\dert \rho + \derx(\rho v) &= 0, \label{hydro_eq1}\\
	\dert v + v \derx v = - \frac{1}{m} \derx V(x,t) &- \frac{1}{m \rho} \derx P(x,t), \label{hydro_eq2}\\
	\dert s + v \derx s &= 0,
	\label{hydro_eq3}
\end{align}
where $V(x,t)$ is the trapping potential (given by Eq.~(\ref{harmonic_trapping_potential} in the present case), and $P(x,t)$ is the local pressure that follows from the thermodynamic equation of the state. 

As the equation of state for the TG gas is the same as that for an ideal Fermi gas, the solutions to Eqs. (\ref{hydro_eq1}) through (\ref{hydro_eq3}) are also the same as those for the ideal Fermi gas; they are given by the following scaling transformations \cite{Bouchoule2016,Atas2017b},
\begin{align}
	\rho(x,t) &= \rho_{0}(x/\lambda(t))/\lambda(t), \label{hydro_rho}\\
	v(x,t) &= x \dot{\lambda}(t) / \lambda(t), \label{hydro_v}\\
	T(t) &= T_{0} / \lambda^{2}(t). \label{hydro_T}
\end{align}
Here, $\rho(x,t=0) = \rho_{0}(x)$ is the initial density profile and $T_{0}$ is the initial temperature of the gas, whereas $\lambda(t)$ is a scaling parameter which satisfies the same Ermakov-Pinney equation (\ref{ermakov}) as in the exact many-body treatment \cite{Bouchoule2016,Atas2017b}. Therefore, under sinusoidal modulation of $\omega(t)$ as in Eq. (\ref{frequency_modulation}) below, one can use the mapping to Mathieu's equation and the same solutions for the scaling parameter $\lambda(t)$ that follow from them. Accordingly, the stability properties of these solutions (see Sec.~\ref{Stablity_of_solutions}) have the same implications on the hydrodynamics of the TG gas as in the exact the many-body treatment, implying that the ensuing stability diagram and the structure of parametric resonances of the TG in a frequency-modulated harmonic trap can be reproduced exactly from the hydrodynamic approach.

Specific examples of calculations of the dynamics of the TG gas that follow from the hydrodynamic approach will be presented in Sec.~\ref{Evolution_of_the_density_and_momentum_distributions}. These calculations are carried out as in Refs.~\cite{Bouchoule2016,Atas2017b} and include the dynamics of the momentum distribution of the gas. The latter does not trivially follow from the above solutions for the real-space density distribution. Despite this, the momentum distribution of the gas can be constructed from the solutions for the density distribution, owing to the local density approximation that is intrinsic to the hydrodynamic approach. For ease of reference we outline this construction in Appendix \ref{Hydro_Theory}.

\section{Mapping to Mathieu's equation for sinusoidal modulation}
\label{Mapping_to_Mathieu_eq}

From here on we focus on the dynamics of the harmonically trapped TG gas in response to a sinusoidal modulation of the trap frequency,
\begin{equation}
\omega^{2}(t)=\omega_{0}^{2}\big{(} 1-\alpha \sin(\Omega t)\big{)}. 
\label{frequency_modulation}
\end{equation} 
Here, $\omega_0$ is the trap frequency in the preparation stage of the initial ($t\leq 0$) thermal equilibrium state, whereas 
$\Omega$ and $\alpha$ ($0\leq \alpha \leq 1$), which define the parameter space of the problem, characterise the frequency and amplitude of subsequent modulation of $\omega(t)$ after time $t>0$.

Using $\omega^{2}(t)$ given by Eq.~(\ref{frequency_modulation}), along with the change of variable $\Omega t= \pi/2-2\tau$, one can show that the homogeneous differential equation (\ref{homoegeneous_ermakov}) takes the canonical form of  Mathieu's equation \cite{Mathieu1868} for the function $z(\tau)=\lambda\big{(}(\pi/2-2\tau)/\Omega \big{)}$,
\begin{equation}
\ddot{z}(\tau)+\big{(} a-2q \cos (2 \tau)\big{)}\,z(\tau)=0, 
\label{Mathieu_eq}
\end{equation}
which is parametrized in terms of 
\begin{equation}
 a=\left(\frac{2 \omega_{0}}{\Omega}\right)^2\text{~~and~~}q=\frac{2\omega_{0}^{2}\alpha}{\Omega^{2}}. \label{parameter_Mathieu}
\end{equation}

The two independent solutions of Mathieu's equation are given by the even and odd Mathieu functions denoted, respectively, by $C(a,q,\tau)$ and $S(a,q,\tau)$, and therefore the two independent solutions to the homogeneous Ermakov-Pinney equation can be written as
\begin{equation}
\lambda_{1}(t)=C\left( \frac{4\omega_{0}^{2}}{\Omega^{2}},\frac{2\alpha \omega_{0}^{2}}{\Omega^{2}},\frac{\pi}{4}-\frac{\Omega t}{2}\right), \label{lambda1}
\end{equation}
\vspace{-0.3cm}
\begin{equation}
\lambda_{2}(t)=S\left( \frac{4\omega_{0}^{2}}{\Omega^{2}},\frac{2\alpha \omega_{0}^{2}}{\Omega^{2}},\frac{\pi}{4}-\frac{\Omega t}{2}\right). \label{lambda2}
\end{equation}

By combining these two solutions and solving the constraint equations for the coefficients $\mathcal{A}$, $\mathcal{B}$ and $\mathcal{C}$ in Eq.~(\ref{general_solution_ermakov}), the general solution to the Ermakov-Pinney equation can be written as
\begin{align}
\notag \lambda(t)=\frac{1}{|W|}&\left[ \omega^{2}(0)\Big{(}\lambda^{\,}_{2}(0)\lambda_{1}(t)-\lambda_{1}(0)\lambda_{2}(t)\Big{)}^{2} \right. \\
& \hspace*{0.5cm}+ \left. \left( \dot{\lambda}_{2}(0)\lambda_{1}(t)-\dot{\lambda}_{1}(0)\lambda_{2}(t)\right)^{2}\right]^{1/2}. 
\label{Scaling_sol}
\end{align}
This function indeed satisfies the initial conditions $\lambda(0)\!=\!1$ and $\dot{\lambda}(0)\!=\!0$, and we note that since the Wronskian of two independent solutions of a differential equation cannot be equal to zero the function $\lambda(t)$ is well defined.

We mention here that physically a slightly different modulation scenario to Eq.~(\ref{frequency_modulation}) is to use cosine modulation of the form 
\begin{equation}
\omega^{2}(t)=\omega_{0}^{2}(1+\alpha \cos(\Omega t)).
\end{equation}
In this case, the modulation of frequency begins slowly and without discontinuity in the derivative (\emph{i.e.}, with a zero slope in the time derivative, whereas in the former case of sine modulation the modulation begins abruptly). Nevertheless, the case of cosine modulation can still be solved using the solutions we present in this work for sine modulation; however, one must be careful in making the appropriate variable transformations. Specifically, the solutions for cosine modulation can be obtained from the solutions (\ref{lambda1}) and (\ref{lambda2}) by a simple time translation $t\rightarrow t-\pi/2 \Omega$ and setting $q\rightarrow-q$, where $q$ is defined in Eq. (\ref{parameter_Mathieu}).

\section{Solutions to Mathieu's equation}
\label{Solutions to Mathieu's equation}

In what follows, we discuss the explicit form of the Mathieu functions, the method of calculating them, and their stability properties in 
the $(a,q)$ parameter space, which ultimately determine the physical properties of the periodically modulated TG gas.
We consider the frequency modulation given by Eq.~(\ref{frequency_modulation}) and restrict ourselves to the case of $q>0$, or equivalently $\alpha>0$ (from Eq. (\ref{parameter_Mathieu})), with the recognition that the transformation $\tau\rightarrow \tau+\pi/2$ simply changes the sign of $q$ in Eq.~(\ref{Mathieu_eq}). Therefore the case $q<0$ ($\alpha<0$) can be treated by simply translating the time $t\rightarrow t+\pi/ \Omega$ in the solutions for $q>0$,(\textit{i.e.}, take $t\rightarrow t+\pi/ \Omega$ with $q>0$ in Eqs. (\ref{lambda1}) and (\ref{lambda2}) and propagate these solutions through Eq. (\ref{Scaling_sol}) to obtain $\lambda(t)$ for the equivalent $q<0$ case).

\subsection{Construction of the two independent solutions to Mathieu's equation}

The differential equation (\ref{Mathieu_eq}) represents the motion of a classical particle in a $\pi$-periodic potential, and therefore from Floquet's theorem we have that a first fundamental solution can be written in the form
\begin{equation}
z_{1}(\tau)=e^{i \nu \tau}f(\tau) \label{fundamental_solution},
\end{equation} 
where $\nu$ is the Floquet exponent (also referred to as the Mathieu characteristic exponent), which depends on the parameters $a$ and $q$, and $f(\tau)$ is a periodic function with period $\pi$, \textit{i.e.}, $f(\tau+\pi)=f(\tau)$. 
By virtue of its $\pi$-periodicity,  the function $f$ can be represented using the Fourier expansion 
\begin{equation}
f(\tau)=\sum_{n=-\infty}^{\infty}c_{n}e^{2in\tau}, \label{Fourier_Expansion}
\end{equation}
where $c_n$ are the expansion coefficients. 

Inserting this form of $f(\tau)$ into Mathieu's equation (\ref{Mathieu_eq}) one obtains a system of linear equations for the coefficients $c_n$, which can be written in matrix form. The condition of having a nontrivial solution to this set of equations results in a matrix determinant equation involving the Floquet exponent $\nu$ (see Appendix \ref{Floquet-exp} for details). For given values of $a$ and $q$ this equation can be solved to obtain $\nu$. In particular, for $a\neq 4j^2$ (where $j$ is an integer) one finds that 
\begin{equation}
	\nu =\frac{2}{\pi}\arcsin \left(\sqrt{\Delta(0)\sin^{2}(\sqrt{a}\pi/2)}\right),
\label{Floquet_Exponent_exact1} 
\end{equation} 
whereas for $a=4j^2$ ($j\in \mathbb{Z}$) one finds that
\begin{equation}
\nu =\frac{1}{\pi}\arccos \big(2\Delta(1)-1\big).
\label{Floquet_Exponent_exact2}
\end{equation}
Here, $\Delta(0)$ and $\Delta(1)$ are numerical coefficients to be obtained from 
\begin{equation}
	\Delta(\nu) \equiv \mathrm{det}\left(\delta_{nm}+\frac{(\delta_{n+1,m}+\delta_{n-1,m})\,q}{(2n+\nu)^2-a}\right),\label{determinant_Delta_1}
\end{equation}
at $\nu=0$ and $\nu=1$, respectively. In the above equation, the integer indices $n$ and $m$ in the infinite determinant run from $-\infty$ to $\infty$, but in practice they must be truncated at a high absolute value that is sufficient for numerical convergence.

From the properties of the matrix determinant involving $\nu$, one can further show (see Appendix \ref{Floquet-exp}) that if $\nu$ is a solution to the determinant equation, then $\pm \nu \pm 2n$ (with $n$ being an integer) is also a solution. Therefore, from Eqs.~(\ref{Floquet_Exponent_exact1}) and (\ref{Floquet_Exponent_exact2}) one can deduce that when the arguments of the inverse trigonometric functions lie within the interval $[-1,1]$ (depending on the values of $a$ and $q$) it is sufficient  to only consider solutions for $\nu$ in the interval $\nu\in [0,1]$. The cases with $\nu=0$ and $\nu=1$ correspond, respectively, to even and odd integer values of the Floquet exponent.
When, on the other hand, the arguments of the inverse trigonometric functions in (\ref{Floquet_Exponent_exact1}) or (\ref{Floquet_Exponent_exact2}) lie outside the interval $[-1,1]$, then the Floquet exponent becomes complex, in which case it can be either a pure imaginary number $\nu=i \eta$ or have the form $\nu=1+i \eta$, with $\eta<0$. 

Once the Floquet exponent $\nu$ is known from Eq.~(\ref{Floquet_Exponent_exact1}) or (\ref{Floquet_Exponent_exact2}), one can solve the system of equations for the Fourier  coefficients $c_n$ (see Appendix \ref{Fourier-coeff}) and finally construct the first fundamental solution $z_1(\tau)$, Eq.~(\ref{fundamental_solution}), in explicit form.
For constructing the second fundamental solution, $z_2(\tau)$, we note that
Mathieu's equation is unchanged by the time reversal transformation $\tau\!\rightarrow\! - \tau$. This means that
\begin{equation}
z_{2}(\tau)=z_{1}(-\tau)=e^{-i \nu \tau}f(-\tau) \label{second_fund_sol}
\end{equation} 
is also an independent solution of Mathieu's equation (\ref{Mathieu_eq}), provided that the Floquet exponent is either complex or real but not an integer. Therefore, the function $z_2(\tau)$ given by Eq. (\ref{second_fund_sol}) can be taken as the second independent solution of Mathieu's equation, in these cases.

When $\nu$ is an integer (we recall that the only two integer values of $\nu$ that we need to consider are $\nu=0,1$), the first fundamental solution $z_{1}(\tau)$ becomes $\pi$ or $2\pi$ periodic (according to the parity of $\nu$), and by Ince's theorem  \cite{mclachlan1964theory} the second linearly independent solution cannot have either a period $\pi$ or $2\pi$. This means that $z_{2}(\tau)$, defined by Eq.~(\ref{second_fund_sol}), no longer constitutes an independent solution of Mathieu's equation (\ref{Mathieu_eq}) as it becomes simply proportional to $z_{1}(\tau)$. Instead, the second independent solution in this case is given by $z_{2}(\tau)=\beta \tau z_{1}(\tau)+g(\tau)$  \cite{mclachlan1964theory,arscott2014periodic}, where $\beta$ is a constant and the function $g(\tau)$ has the same periodicity as $z_{1}(\tau)$. An explicit expression for $z_2(\tau)$ in terms of a series of products of Bessel functions is given in Appendix \ref{Fourier-coeff}.
 
In conclusion, except for the special case when $\nu=0$ or $1$, the two independent solutions $z_{1}(\tau)$ and $z_{2}(\tau)$ can always be combined to form the even and odd solutions of Mathieu's equation,
\begin{align}
C(a,q,\tau)&=\frac{z_{1}(\tau)+z_{2}(\tau)}{2z_{1}(0)}, \label{Cosine_elliptic}\\ 
S(a,q,\tau)&=\frac{z_{1}(\tau)-z_{2}(\tau)}{2\dot{z}_{1}(0)},\label{Sine_elliptic}
\end{align}
such that $C(a,q,0)=1$ and $\dot{C}(a,q,0)=0$, whereas $S(a,q,0)=0$ and $\dot{S}(a,q,0)=1$. Note that the normalizations used here is quite arbitrary and has no consequences in the construction of the scaling function $\lambda(t)$ as the normalization constants are absorbed by the coefficients $\mathcal{A}$, $\mathcal{B}$ and $\mathcal{C}$ in Eq.~(\ref{general_solution_ermakov}).

\subsection{Stability of solutions to Mathieu's equation} \label{Stablity_of_solutions}

The long-time asymptotic behaviour of the scaling function $\lambda(t)$, and hence the behaviour of physical observables such as the density and momentum distributions of the trapped TG gas, are in direct bijection with the stability properties of the solutions to Mathieu's equation. As such, to characterise the long-time dynamical behaviour of the harmonically trapped TG gas, one need only to examine the stability properties of these solutions.
The solutions are said to be stable if they remain bounded when $\tau\rightarrow \infty$, and they are unstable if they tend to $\pm \infty$ when $\tau\rightarrow \infty$.  From Eqs.~(\ref{fundamental_solution}) and (\ref{second_fund_sol}), we can see see that stable solutions correspond to $\nu$ being real, whereas unstable ones correspond to $\nu$ being complex. In addition to their stability, the solutions can be classified according to their periodicity:
when $\nu=p/r$ is a rational fraction less than unity, with $p$ and $r$ being mutually prime integers, the solutions have a period of $2\pi r$, and when $\nu$ is an irrational number the solutions have no specific periodicity.

\begin{figure}[tbp]
\includegraphics[width=7.5cm]{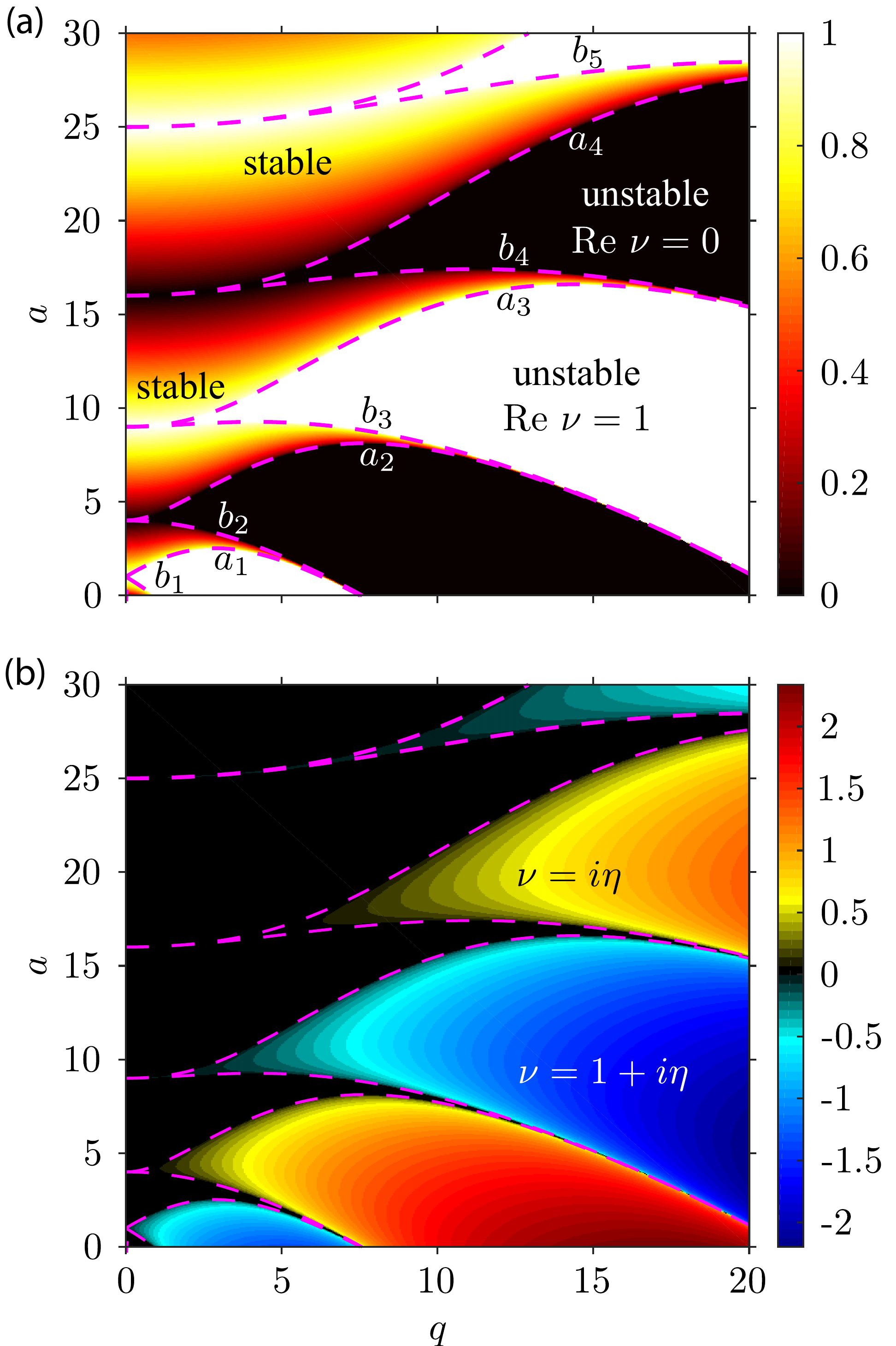}
\caption{Real (a) and imaginary (b) parts of the Floquet exponent $\nu$ in the $(a,q)$ plane. The dashed lines separate the regions in which the solutions to Mathieu's equation are stable [colormapped regions in (a), or equivalently black regions in (b)] and unstable [black and white regions in (a), or  the colormapped regions in (b)]; see text for details. } \label{Figure1}
\end{figure}

In Figs.~\ref{Figure1}\,(a) and (b) we show the density plots of, respectively, the real and imaginary parts of the Floquet exponent $\nu$ in the $(a,q)$ plane, calculated using Eqs.~(\ref{Floquet_Exponent_exact1}) and (\ref{Floquet_Exponent_exact2}). On these plots, the dashed (magenta) lines show the boundaries between the stable and unstable regions. The Floquet (or the Mathieu characteristic) exponent on these boundaries is an integer (modulo the periodicity), $\nu(a,q)=n\, \mathrm{mod}\, 2$ (\textit{i.e.}, $\nu = 0$ or $1$), with $n\in \mathbb{Z}$ indexing the band. For a given $q$, the values of the parameter $a$, for which the Floquet exponent $\nu(a,q)$ lies on the stability boundaries are called the characteristic values of Mathieu's equation and are denoted as $a_{n}(q)$ for even (cosine elliptic) solutions  and $b_{n}(q)$ for odd (sine elliptic) solutions.
As can be seen from the Floquet solution (\ref{fundamental_solution}) and the Fourier expansion  (\ref{Fourier_Expansion}), on the even band 
 the fundamental solution $z_1(\tau)$ is $\pi$ periodic, whereas on the odd band the fundamental solution has period $2\pi$. We recall that, for $\nu = 0$ or $1$, only the fundamental solution $z_1(\tau)$ has a period of $\pi$ or $2\pi$, whereas the second independent solution is not periodic and is given by a series involving products of Bessel functions (see Appendix \ref{Fourier-coeff}).

The parameter space between the different bands corresponding to unstable solutions is shown in Fig.~\ref{Figure1}\,(a) by black (where $\mathrm{Re} \, \nu=0$) and white (where $\mathrm{Re} \, \nu=1$) regions. On the other hand, the parameter space corresponding to stable solutions is shown by the colour-mapped regions, located in $a_{n}(q)<a(q)<b_{n+1}(q)$ (with $n=0,1,2,\dots$ and $a_0<0$). The Floquet exponent is real in these regions, with $0<\nu<1$, and the solutions to Mathieu's equation are given by
(\ref{fundamental_solution}) and (\ref{second_fund_sol}). The Fourier expansion coefficients in Eq.~(\ref{Fourier_Expansion}) are also real in these regions, and lead to real-valued solutions.

Respectively, in Fig.~\ref{Figure1}\,(b), the black regions, where $\mathrm{Im}\, \nu=0$, correspond to stable solutions, whereas the colour-mapped regions  correspond to unstable solutions. In the unstable regions, the Floquet exponent is either pure imaginary $\nu=i\eta$, with $\eta>0$, which is the case in the regions $ b_{2n}(q)<a(q)<a_{2n}(q)$ [where $\mathrm{Re} \, \nu=0$ in Fig.~\ref{Figure1}\,(a)], or it is given by  
$\nu=1 + i\eta$, with $\eta<0$, which is the case in the regions $ b_{2n+1}(q)<a(q)<a_{2n+1}(q)$ [where $\mathrm{Re} \, \nu=1$ in Fig.~\ref{Figure1}\,(a)].
In these unstable regions, the Fourier expansion coefficients are generally complex, which ultimately leads to complex solutions of Mathieu's equation. In practice, however, it is simpler to work with real valued solutions, and in Appendix \ref{Fourier-coeff} we show how one can use the symmetry of the Fourier expansion coefficients in order to absorb their complex phase factor in the definition of the first and second fundamental solutions, after which these solutions become real valued.

\begin{figure}[tbp]
	\includegraphics[width=6.4cm]{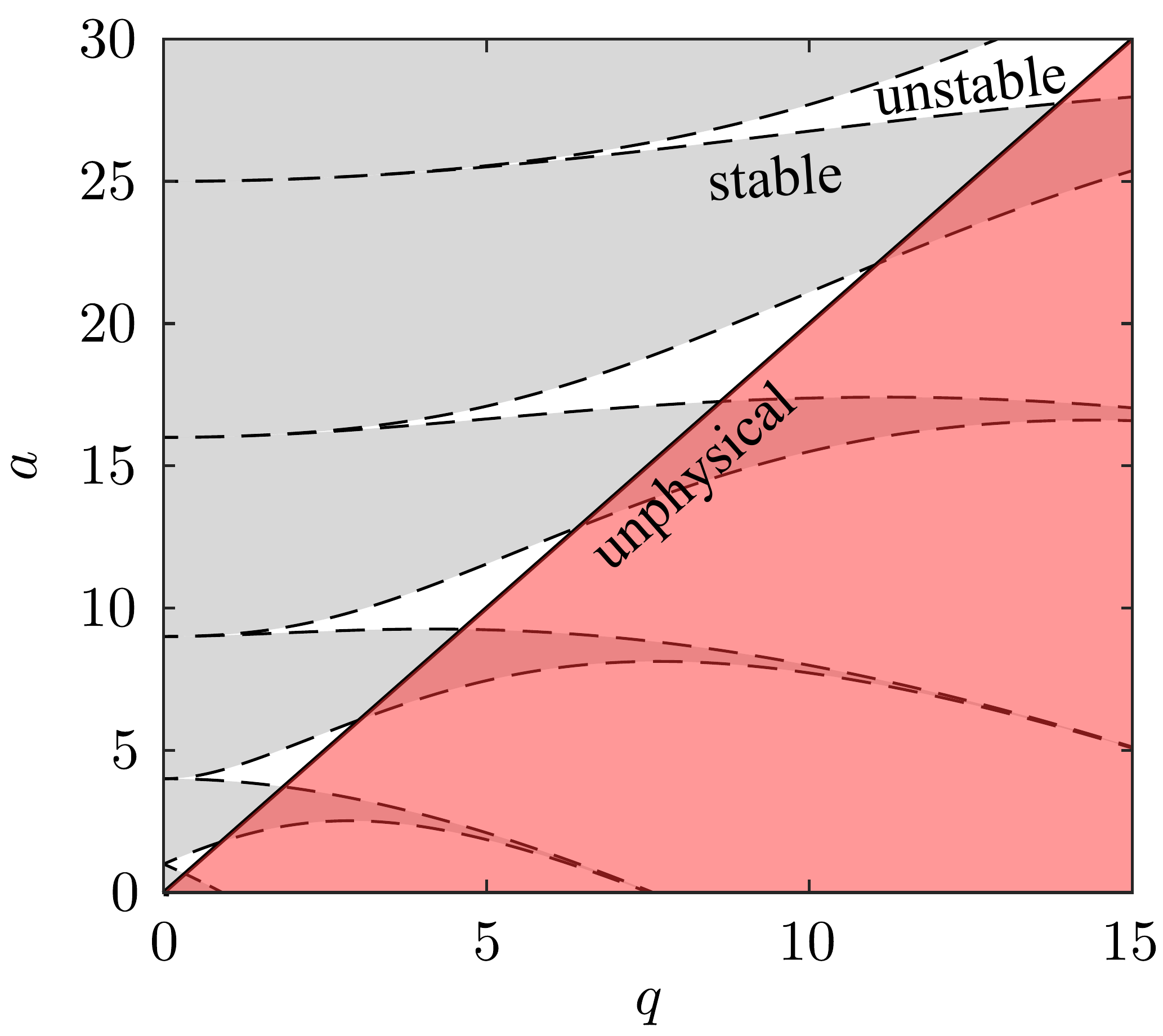}
	\caption{Stability diagram of solutions to Mathieu's equation in the $(a,q)$ parameter space. The grey regions correspond to stable solutions, whereas the white regions to unstable solutions. The red shaded region where $a< 2q$ corresponds to physically unattainable parameter space, which follows from the restriction $0\leq \alpha \leq 1$ on the modulation amplitude in Eq.~(\ref{frequency_modulation}) and from the fact that the pair $(a,q)$ are not independent, but are related by Eq.~(\ref{Mathieu_Eq_parameters}) (see text).}
	\label{Figure2}
\end{figure}

Discarding the actual values of the Floquet exponent in the $(a,q)$ parameter space and concentrating merely on whether $\nu$ is real or complex leads to the familiar stability digram (see, \textit{e.g.}, \cite{mclachlan1964theory}) of solutions to Mathieu's equation, which is shown in Fig.~\ref{Figure2} in the positive quadrant of the $(a,q)$ plane.

\section{Stability diagram of the TG gas} \label{Stability_diagram_of_the_TG_gas}

We now return to our physical problem of the TG gas in the frequency-modulated harmonic trap. The frequency modulation is parametrised in terms of two independent parameters, $\Omega$ and $\alpha$, characterising, respectively, the frequency and amplitude of modulation of the trap frequency through Eq.~(\ref{frequency_modulation}). Mathieu's equation, on the other hand, is parametrised in terms of $a$ and $q$, and the mapping between these pairs of parameters is given by Eq.~(\ref{parameter_Mathieu}). This particular form of mapping means that Mathieu's parameters $a$ and $q$ are not independent, but are related by
\begin{equation}
 	a=\left(\frac{2 \omega_{0}}{\Omega}\right)^2, \quad q=\frac{\alpha}{2} a. \label{Mathieu_Eq_parameters}
\end{equation}
Moreover, the physical restriction $0\leq \alpha \leq 1$ on the amplitude parameter $\alpha$ restricts the values of $q$ to $0\leq q \leq a/2$, hence the physically unattainable region of $q>a/2$ or $a<2q$, shown in Fig.~\ref{Figure2}.

\begin{figure}[tbp]
\includegraphics[width=6.4cm]{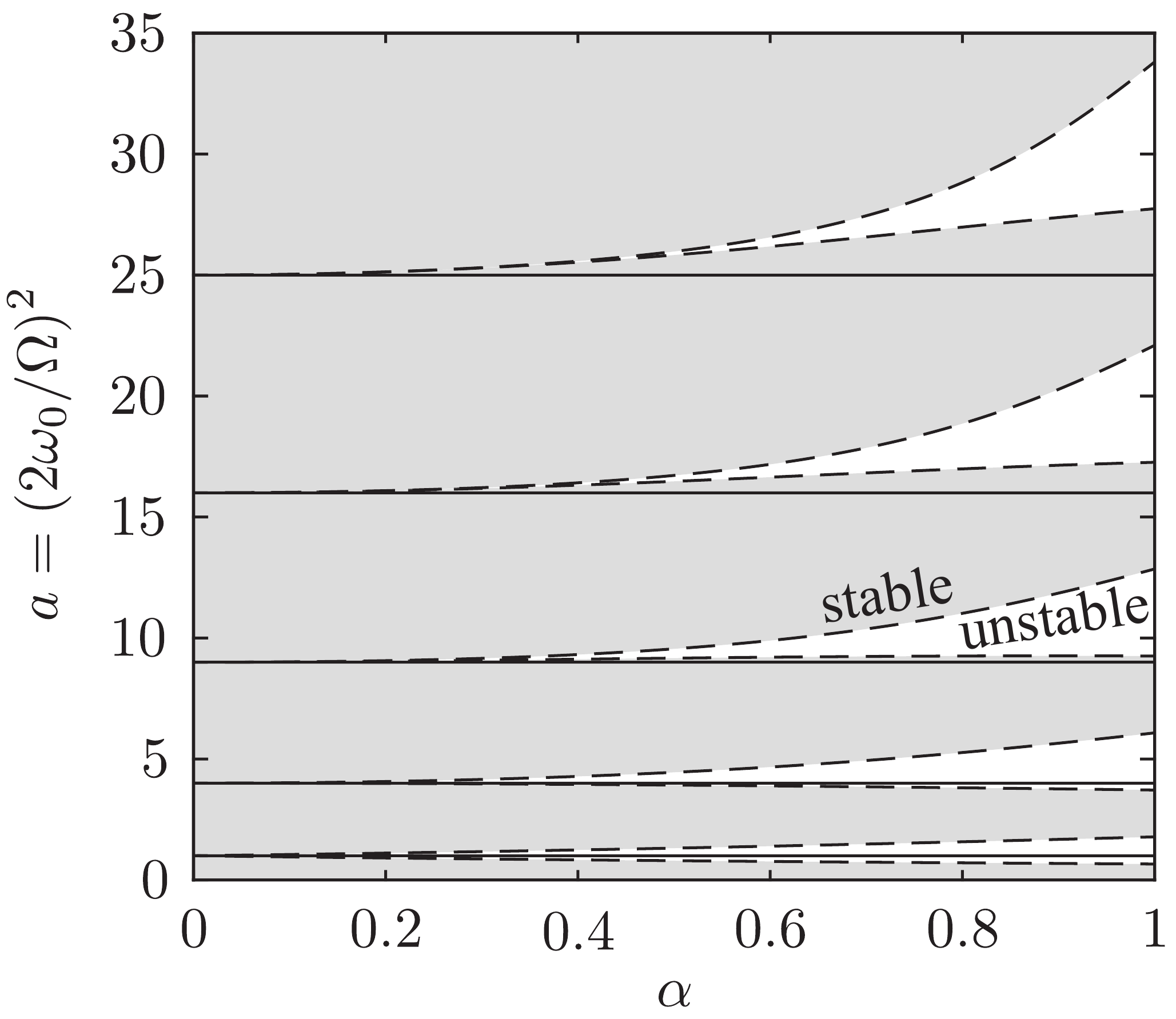}
\caption{Stability diagram of the dynamics of the TG gas. Grey regions correspond to stable dynamics, whereas the white regions to unstable dynamics. Straight horizontal lines at $a=1,4,9,16,25$ correspond to a set of fractional driving frequencies $\Omega_{j}=2\omega_{0}/j$ (with $j=1,2,3,4,5$), which lead to parametric resonances for small driving amplitudes $\alpha$ (see text for further details). }
\label{Figure3}
\end{figure}

For given values of the physical parameters $(\Omega,\alpha) $ (where $\Omega$ is measured in units of $\omega_0$), the point $(a,q)$ is uniquely specified, and gives the stability of the solution to Mathieu's equation through the Floquet exponent $\nu(a,q)$. However, from an experimental point of view it is more practical to fix, for instance, the modulation amplitude $\alpha$ and search for modulation frequencies $\Omega$ (or values of $a$) that lead to the desired stable or unstable dynamics. It is therefore more convenient to work in the $(a,\alpha)$ parameters space, rather than in the $(a,q)$ space.

Using $(a,\alpha)$ as our parameters, we can now map the stability diagram of Fig.~\ref{Figure2} to the parameter space of $(a,\alpha)$ by first fixing the amplitude $\alpha$, or equivalently the slope $2/\alpha$ of any straight line $a=(2/\alpha)q$ in Fig.~\ref{Figure2}, and then following the trace of this line through stable and unstable regions. The intersections of this line with consecutive characteristic lines $a_{n}(q)$ and $b_{n+1}(q)$, where $n=0,1,2,\dots$, define a set of values of $q$ and hence $a$ in between which the solution is stable.  In this way, we can arrive at the new stability diagram in the $(a,\alpha)$ parameter space shown in Fig.~\ref{Figure3}, which is equivalent to the physically attainable part of the diagram of Fig.~\ref{Figure2}.

\subsection{Parametric Resonances} 
\label{Parametric_Resonances}

The stability  diagram of Fig.~\ref{Figure3} can also be read in terms of an experimental scenario in which one fixes the modulation frequency $\Omega$ (or the parameter $a$) and then searches for the modulation amplitude $\alpha$ for stable or unstable dynamics. This corresponds to following a straight horizontal line at a fixed value of $a$ in Fig.~\ref{Figure3}; the values of $\alpha$ for which such a line remains in the white (grey) region give the range of modulation amplitudes for which the dynamics are unstable (stable).
Examples of such lines, corresponding to driving frequencies $\Omega$ which are equal to fractions of $2\omega_{0}$,
\begin{equation}
	\Omega_{j}=2\omega_{0}/j, \label{two_level_resonance}
\end{equation}
are shown in Fig.~\ref{Figure3}\,(a) for $j=1,2,..,5$. 

These special values of frequencies are the same as in the phenomenon of parametric resonance, well-known in the problem of a classical oscillator subject to (an externally driven) time-variation of the parameters \cite{landau1976mechanics}, as well as in the problem of a (single-particle) quantum oscillator with time-varying frequency \cite{popov1969parametric}. In these systems, as well as in the present many-body problem of the TG gas, the oscillatory response of the system can be resonantly amplified (and is hence unstable) at certain driving frequencies. 

The primary parametric resonance  corresponds to $a=1$, or to $j=1$ in $\Omega_{j}=2\omega_{0}/j$, which is the frequency of the natural breathing mode oscillations \cite{Atas2017b} of the TG gas. It occurs for arbitrary values of the modulation amplitude $\alpha$ as indicated by the fact that the respective horizontal line in Fig.~\ref{Figure3}\,(a) remains in the white (unstable) region for all $0 < \alpha \leq 1$. The second resonance, corresponding to $a=4$ (or $j=2$), displays a similar behaviour. 

In contrast, for $j \ge 3$ the lines of fixed $a=j^2$ (or $a= 9, 16, 25, ...$) start (for $\alpha > 0$) in the gray (stable) regions. This is because the lower boundaries of the unstable regions (dashed lines in Fig.~\ref{Figure3}) corresponding to $j \geq 3$ all have positive curvature at small $\alpha$ and remain above the respective values of $a=9, 16, 25, ...$ for all $\alpha$. As such, the resonant enhancement (\textit{i.e.}, unstable behavior) occurs at values of $a$ that are shifted upwards relative to $a=9, 16, 25, ...$ (in terms of frequencies, $\Omega$ is shifted downward relative to the exact fractional values of $\Omega_{j}=2\omega_{0}/j$). We note that this upward shift, as well as the width of unstable regions, increases with $j$; in fact, for relatively small modulation amplitudes, the width of the resonance region $\Delta a_j$ scales as $\Delta a_j \approx \frac{ j^{2j} \alpha^j }{ 2^{3(j-1)} [(j-1)!]^2}$ \cite{bell1957note}. Moreover, for large enough $j$ (namely, $j \geq 8$) it is possible for the 
upper bound of the unstable region to cross the next resonant line of constant $a=64,8 1,...$, or in other words, it becomes possible  
for the horizontal line of fixed $a=64, 81,...$ to re-enter the white (unstable) region of the previous (lower $j$) resonance as $\alpha$ is increased from 0 to 1.


The structure of resonances of the parametrically driven TG gas has been previously discussed in Ref.~\cite{QuinnHaque2014}, however, 
the exact boundaries of stable and unstable regions, as obtained here in Fig.~\ref{Figure3} from the stability properties of Mathieu's equation, have not been identified to the best of our knowledge.

\section{Evolution of the density and momentum distributions} 
\label{Evolution_of_the_density_and_momentum_distributions}

We first reiterate that the long-time dynamical behaviour of the trapped TG gas can be determined through the stability diagram presented in Fig.~\ref{Figure3}. For analysing the transient behaviour, however, one needs to resort to explicit time-dependent calculations of the dynamics. Such calculations provide insight on how the choice of modulation parameters, as well as the finite temperature of the gas, can affect  dynamical features at short to intermediate times. In what follows, we thus 
examine both the real-space density and momentum distributions of the gas and present two typical examples which exemplify common behaviours in both the stable and unstable regions of the parameter space, using both the exact many-body approach and the hydrodynamic approach outlined in Sec.~\ref{The_model_and_scaling_solution}. We also show how finite-temperature effects alter the behaviour of the gas in these two examples.

\subsection{Unstable dynamics} \label{Unstable_dynamics}

\begin{figure}[tbp]
	\includegraphics[width=6.4cm]{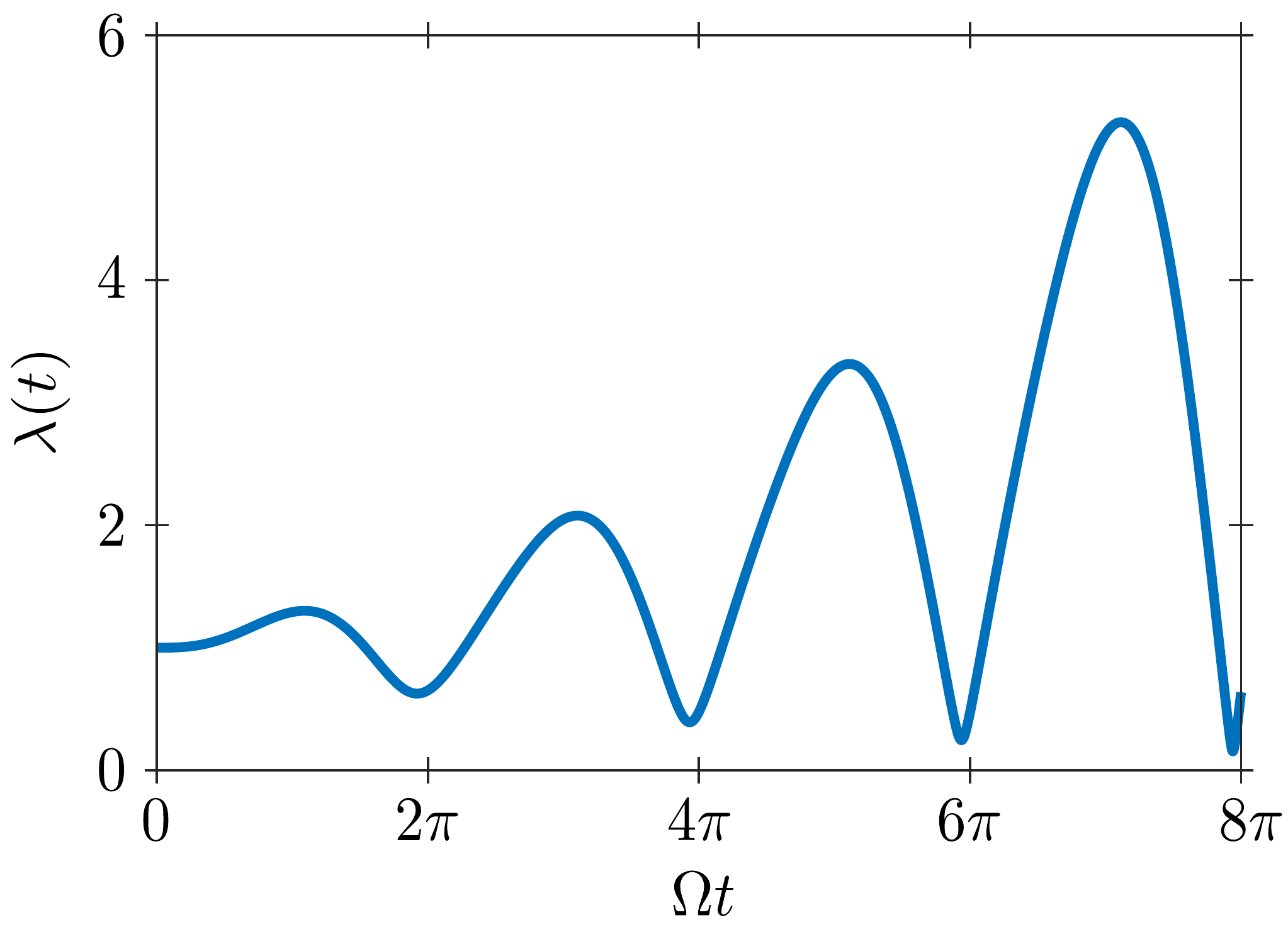}
	\caption{The scaling solution $\lambda(t)$ as a function of the dimensionless time $\Omega t$, for $a=1$ (or $\Omega=2\omega_{0}$) and $\alpha=0.6$.}
	\label{lambda_figure_a}
\end{figure}

In Fig.~\ref{lambda_figure_a} we show the evolution of the scaling function $\lambda(t)$ for modulation parameters $(a,\alpha) = (1,0.6)$. The choice $(a,\alpha) = (1,0.6)$ lies on the primary parametric resonance, corresponding to $j=1$ in Eq.~(\ref{two_level_resonance}) and hence the driving frequency of $\Omega = 2\omega_{0}$, and therefore the dynamics are unstable. Accordingly, the sequence of the peaks in the scaling solution $\lambda(t)$ grows exponentially (on a long-time scale), whereas the oscillations that accompany the exponential growth are due to the natural breathing mode behaviour \cite{Atas2017b}.

\begin{figure*}[btp]
	\includegraphics[width=16cm]{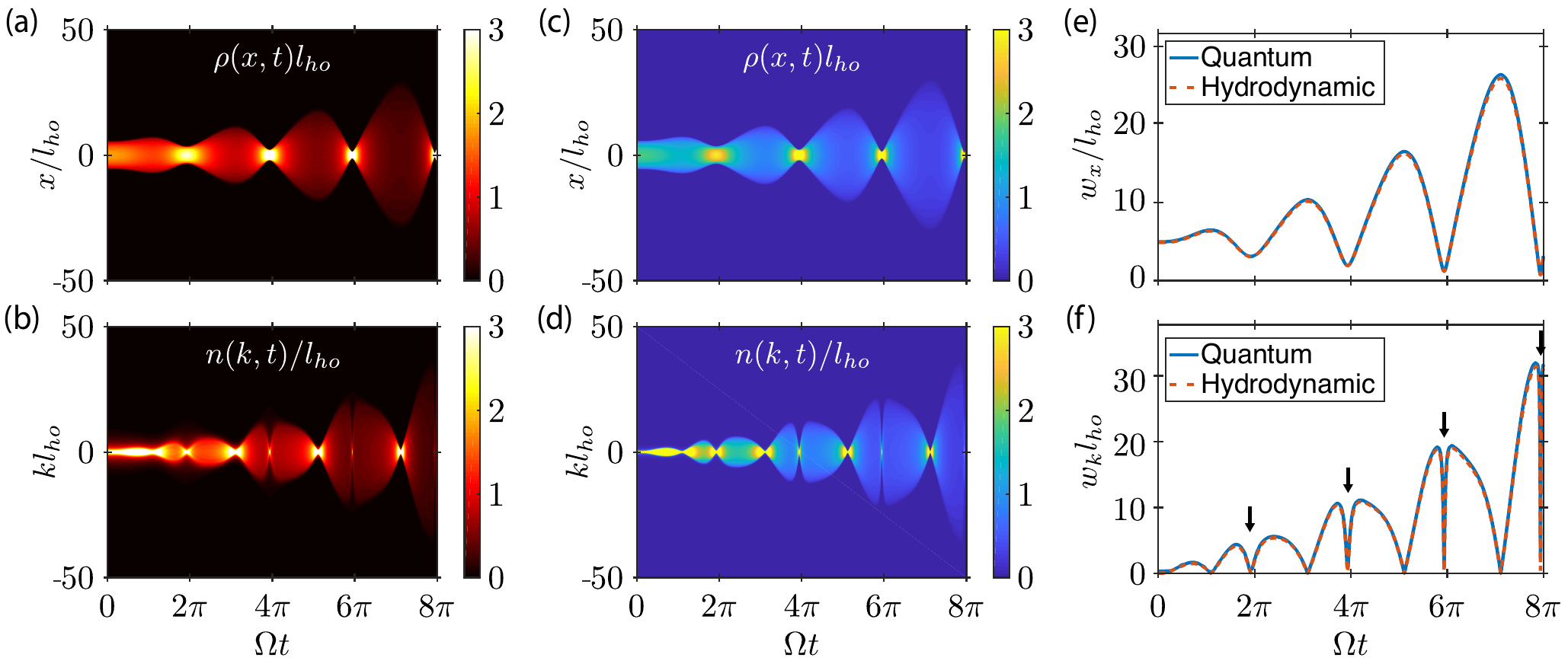}
	\caption{Dynamics of a TG gas containing $N=16$ particles, for $(a,\alpha) = (1,0.6)$ and an initial dimensionless temperature of $\theta_{0} = k_{B} T_{0}/N\hbar\omega_{0} = 0.01$. The dimensionless density $\rho(x,t)l_{ho}$ and  momentum distributions $n(k,t)/l_{ho}$ (where $l_{ho} = \sqrt{\hbar/m\omega_{0}}$ is the harmonic oscillator length) computed using the exact quantum model are shown, respectively, in (a) and (b), whereas the same quantities computed using the hydrodynamic approach are shown in (c) and (d). The dimensionless half-width at half-maximum (HWHM) of the density distribution, $w_{x}(t)/l_{ho}$, as predicted by both theories is shown in (e); similarly, the dimensionless HWHM of the momentum distribution, $w_{k}(t) l_{ho}$, is shown in (f). Instances of the many-body bounce are shown in (f) with arrows. The colour scale for the full height of the distributions $\rho(x,t)$ and $n(k,t)$ is cropped at some point to allow details of the width of these distributions to be seen more clearly. } \label{Unstable_lowT}
\end{figure*}

\begin{figure*}[btp]
	\includegraphics[width=16cm]{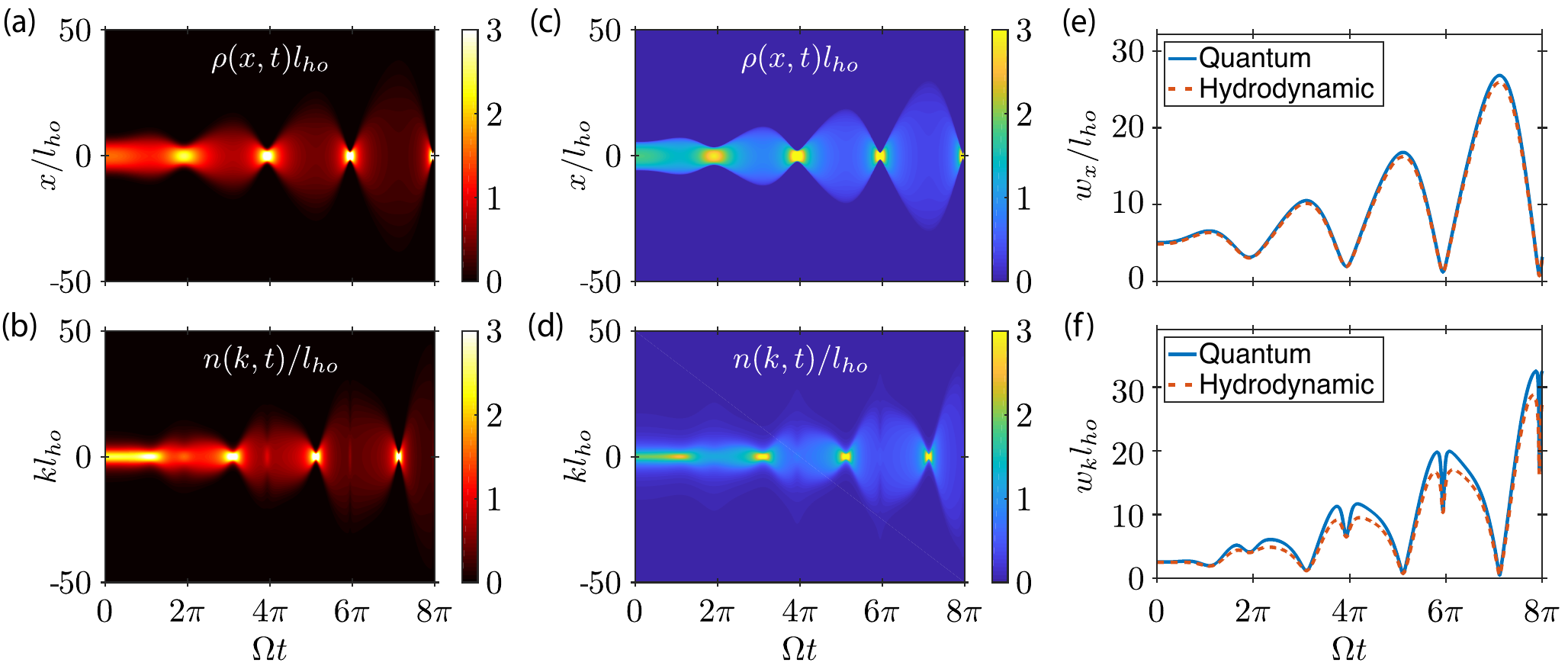}
	\caption{Same as in Fig.~\ref{Unstable_lowT}, but for a higher temperature, $\theta_{0} = k_{B} T_{0}/N\hbar\omega_{0} = 0.5$. }
	\label{Unstable_highT}
\end{figure*}

Figure~\ref{Unstable_lowT} shows the resulting dynamics of a finite temperature TG gas containing $N=16$ particles at a relatively low initial temperature $T_0$. The temperature is parametrized in terms of the dimensionless parameter $\theta_{0} = k_{B} T_{0}/N\hbar\omega_{0} $, where $N\hbar \omega_{0}/k_B$ is the temperature of quantum degeneracy. We show the evolution of the particle number density $\rho(x,t)$, the momentum distribution $n(k,t)$, and the respective widths of the gas determined from both the exact quantum theory and the hydrodynamic approach. According to Eq. (\ref{eq.ss}), the magnitude of the density $\rho(x,t)$ depends inversely on $\lambda(t)$. Thus, the sequential minima of $\rho(0,t)$, when the the density profiles are the broadest, diminish as $\lambda(t)$ grows, whereas the respective peak widths (when the density profile is the narrowest) grow in proportion to $\lambda(t)$. The behaviour of the momentum distribution $n(k,t)$ is reciprocal to that of the real-space density. However, the sequence of narrow and broad distributions is interrupted by an additional narrowing of the momentum distribution at time instances when the density distribution is also narrow. This is a manifestation of the phenomenon of quantum many-body bounce, studied for the breathing mode oscillations of the Tonk-Girardeau gas in Ref.~\cite{Atas2017b}, which in turn is similar to the phenomenon of frequency doubling in the weakly interacting 1D quasicondensate~\cite{Fang2014,Bouchoule2016}.

We note that, for the lower temperature dynamics which we present in Fig.~\ref{Unstable_lowT}, the hydrodynamic theory predicts very well the features obtained from the exact theory.

In Fig.~\ref{Unstable_highT}, we present the dynamics resulting from the same modulation parameters as in Fig.~\ref{Unstable_lowT}, but for a gas at a higher initial temperature of $\theta_{0} = 0.5$. As can be seen from the width of the momentum distribution, the higher temperature example displays thermal broadening. The thermal broadening, however, is apparent only for a short amount of time, and eventually it is overwhelmed by the large-scale dynamical broadening resulting from the unstable nature of the scaling parameter $\lambda(t)$. Accordingly, most of the dynamical features remain largely unaltered by the temperature of the gas at longer times. However, the thermal broadening does significantly affect the many-body bounce \cite{Atas2017b}, which is far less pronounced compared to the low temperature case. 

Within the hydrodynamic theory, which is derived here for temperatures lower than the temperature of quantum degeneracy (so that the density profile is well approximated by an inverted semicircle), the real space density does not depend on temperature. As such, the density profile cannot capture any alterations from thermal effects. On the other hand, the momentum distribution in the hydrodynamic theory is explicitly dependent on temperature through the width of the Lorentzian distribution (see Appendix \ref{Hydro_Theory}). In our higher temperature example, this is evident in the overall broadening of the momentum distribution, which is in agreement with the exact theory. The hydrodynamic theory also accurately captures the blurring effect of the increased temperature on the momentum width at the inner turning points where the many-body bounce occurs, even though the quantitative agreement with the exact theory is poorer in the immediate vicinity of these inner turning points \cite{Atas2017b}.

\subsection{Stable dynamics} \label{Stable_dynamics}

\begin{figure}[tbp]
	\includegraphics[width=6.4cm]{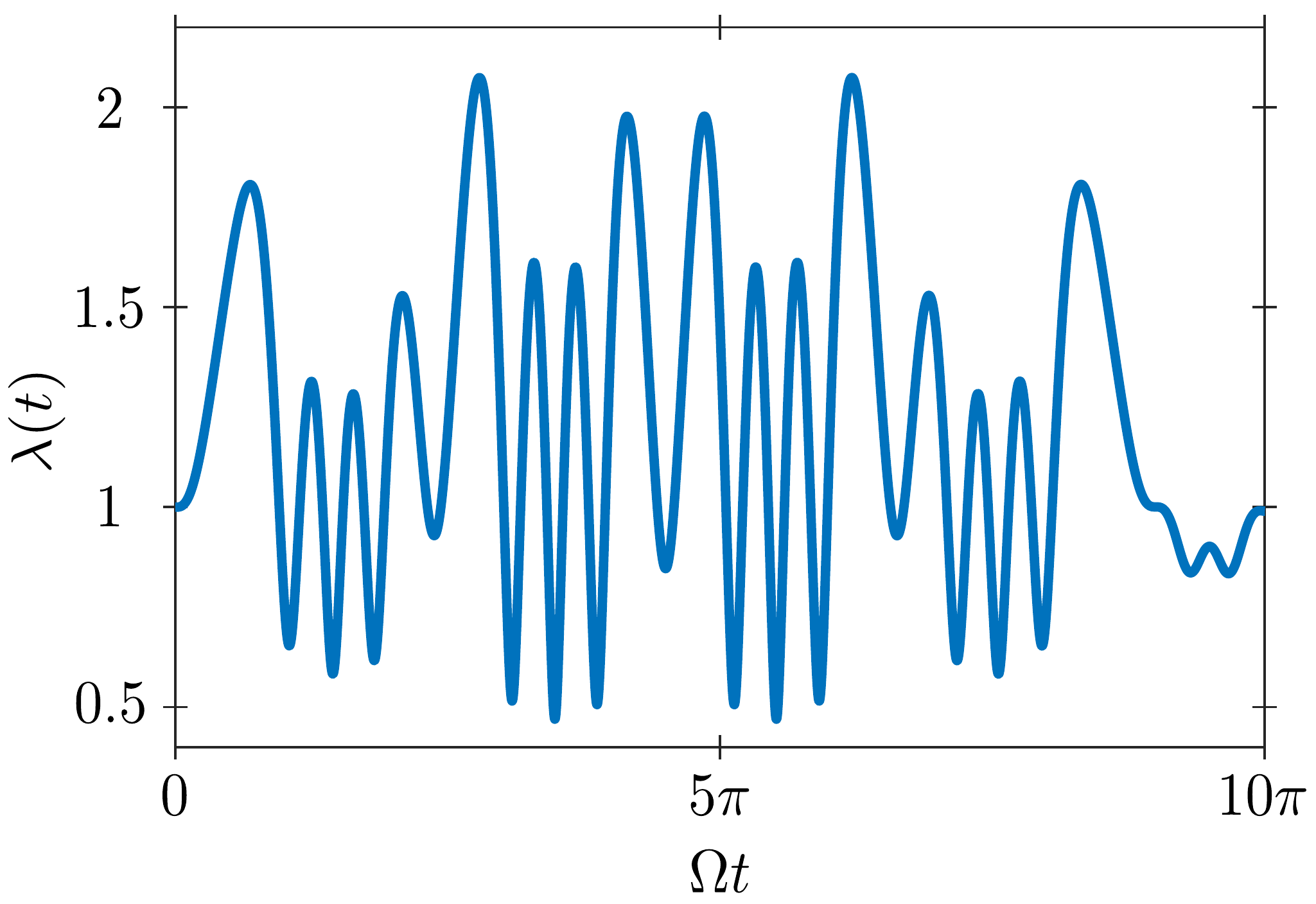}
	\caption{The scaling solution $\lambda(t)$ as a function of the dimensionless time $\Omega t$, for $a=16$ (or $\Omega=\omega_{0}/2$) and $\alpha = 0.8$.}
	 \label{lambda_figure_b}
\end{figure}

In Fig.~\ref{lambda_figure_b} we show the evolution of the scaling function $\lambda(t)$, for modulation parameters $a=16$ (or $\Omega=\omega_{0}/2$) and $\alpha = 0.8$. This parameter choice corresponds to stable or bounded dynamics, which generally can be periodic or aperiodic. More specifically, the Floquet exponent in this example is $\nu \approx 0.1991$, for which the solution is aperiodic. However, this value of $\nu$ is close to $\nu = 1/5$, which results in strictly periodic dynamics (with a period of $10\pi$, following from the periodicity conditions given in Section \ref{Stablity_of_solutions}); accordingly, the example under consideration is nearly periodic and hence we only show the behaviour of $\lambda(t)$ within $\Omega t\in [0,10\pi]$.

\begin{figure*}[tbp]
	\includegraphics[width=16cm]{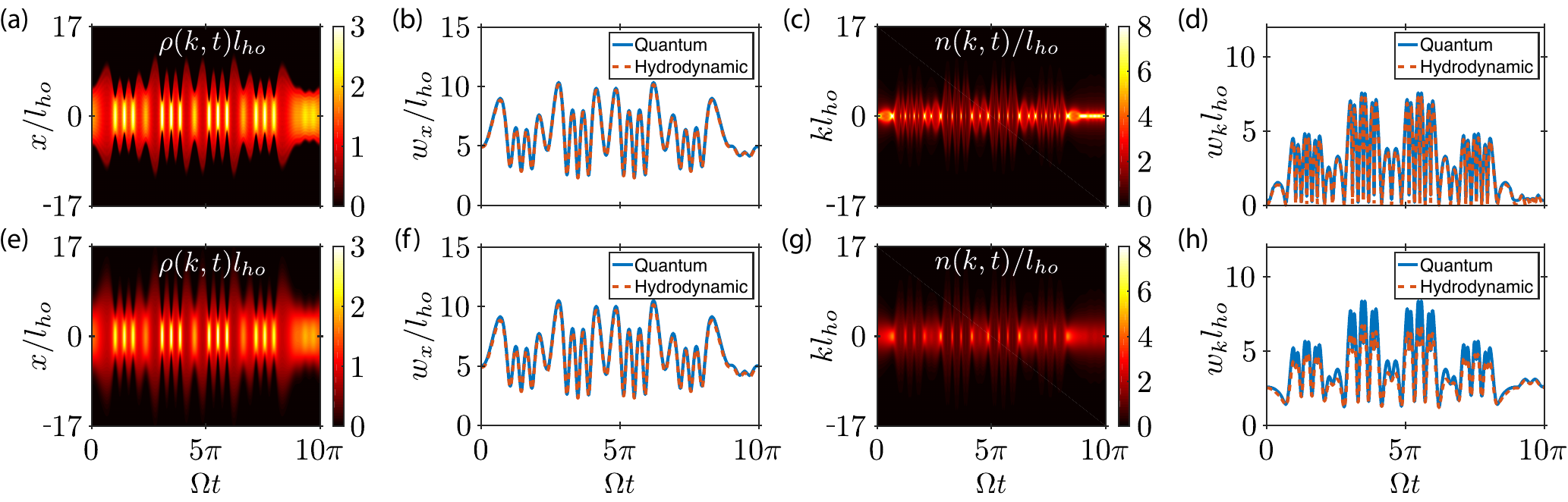}
	\caption{Same as in Figs.~\ref{Unstable_lowT} and \ref{Unstable_highT}, except for modulation parameters from the stable region, $(a,\alpha) = (16,0.8)$. The top and bottom rows correspond, respectively, to an initial dimensionless temperature of $\theta_{0} = 0.01$ and $0.5$, and we no longer show the density and momentum distributions from the hydrodynamic theory.
}
\label{Stable_lowhighT}
\end{figure*}

The high-frequency oscillations in $\lambda(t)$ within this period are a result of beating arising from the contribution of  only a few most significant Fourier coefficients in the expansion of Eq.~(\ref{Fourier_Expansion}). As $\rho(x,t)=\rho(x,0)/\lambda(t)$, the same high-frequency oscillations show up in the dynamics of the density profile and its width shown in Fig.~\ref{Stable_lowhighT}. In contrast, the dynamics of the momentum distribution and its width show oscillations that occur at approximately twice the frequency of oscillations of the density profile due to the phenomenon of quantum many-body bounce.  The thermal broadening effect in the higher temperature example in Fig.~\ref{Stable_lowhighT} is qualitatively the same as in the example of unstable dynamics (Sec.~\ref{Unstable_dynamics}) and is more clearly seen in the width of the momentum distribution of the gas. Similarly, the comparison between the hydrodynamic and exact theories is qualitatively the same as before, and we restrict ourselves to showing the hydrodynamic results only for the widths of the density and momentum distributions [dashed lines in Fig.~\ref{Stable_lowhighT}\,(b), (d), (f), and (h)].

\section{Summary}
\label{Summary}

In conclusion, we have characterised the out-of-equilibrium dynamics of a finite-temperature harmonically trapped TG gas in response to periodic modulation of the trap frequency. The analysis was performed using exact quantum many-body theory as well as a finite-temperature hydrodynamic approach. Due to the scaling transformations, which exist for the single-particle orbitals of the system as well as in the hydrodynamic theory, this otherwise complex many-body problem can be reduced to the solution of a single differential equation, Eq. (\ref{ermakov}). For periodic modulations of the form given by Eq. (\ref{frequency_modulation}), we have shown that the solution to Eq. (\ref{ermakov}) can be mapped to the solutions of Mathieu's differential equation. 

We have provided a detailed analysis of how to construct the solutions to Mathieu's equation and identified the physical and unphysical parameter spaces of these solutions when they are mapped to describe the TG many-body problem. Using the stability properties of Mathieu's equation in terms of Floquet theory, we have constructed the stability diagram of the TG system as a function of the modulation frequency and amplitude. The stability diagrams following from the exact quantum theory and the hydrodynamic approach are identical. We have also identified the structure of parametric resonances within this stability diagram.

In order to explore the transient dynamics of the TG system, we have provided examples of the modulation-induced dynamics in both the unstable and stable regions of the parameter space. We find that the collective many-body bounce effect reported in Ref.~\cite{Atas2017b} persists in both of these dynamical regimes. Furthermore, in each regime we include examples at both a low and high temperature so that we can comment on thermal effects within this system. We note that higher temperatures cause significant thermal broadening to the width of the momentum distribution and also cause blurring of the collective many-body bounce. In the unstable regime, however, the overall thermal broadening of the momentum distribution is eventually suppressed and drowned out by the exponentially increasing nature of the dynamics.

Our results open the path to analysing periodic modulation protocols in the context of nonequilibrium thermodynamics of interacting quantum many-body systems. As an example, they provide the tools for calculating quantum work distributions and designing periodic cycles suitable for quantum many-body heat engines.

In view of applicability of our results to realistic physical systems, we note that the TG model corresponds to the limit of a trapped 1D Bose gas, or the Lieb-Liniger model \cite{Lieb-Liniger1963}, with a two-body contact interaction potential, $g\delta(x-x')$, in which the 1D coupling strength $g\rightarrow \infty$. The relevant dimensionless interaction parameter in the Lieb-Liniger model is $\gamma(x)=mg/\hbar^2\rho(x)$, where $\rho(x)$ is the local 1D density \cite{kheruntsyan2005}. In the TG limit, this local dimensionless interaction parameter must tend to infinity at all $x$. Physically realisable strongly interacting 1D Bose gases are characterised, however, by a very large ($\gamma(x)\gg 1$) but finite interaction strength \cite{paredes2004tonks,Kinoshita2004,Kinoshita2006,Naagerl2009,Naagerl2015}.
Accordingly, the results following from the strictly TG model serve only as a good approximation for such strongly interacting gases. As such, our dynamical results in the unstable regime of modulation (where the peaks of the density $\rho(x,t)$ grow exponentially with time) will eventually become physically invalid as the dimensionless local interaction strength $\gamma(x,t)$ can become of the order of, or much smaller than, unity, for which the TG model becomes inapplicable.

\begin{acknowledgments}
The authors acknowledge stimulating discussions with D.~M.~Gangardt and 
support by the Australian Research 
 Council Discovery Project grant DP170101423.
\end{acknowledgments}

\appendix

\section{Finite-temperature hydrodynamics} 
\label{Hydro_Theory}

For specific dynamical calculations using the scaling solutions to the hydrodynamic equations, the initial density profile in Eq.~(\ref{hydro_rho}) can be approximated by the Thomas-Fermi (TF) semicircle \cite{Atas2017b},
\begin{align}
	\rho_{0}(x) &= \rho_{0}(0) \sqrt{1 - \frac{x^{2}}{R_{\text{TF}}^{2}}}, \label{rho_{0}_TF}
\end{align}
which is valid when the thermal energy scale is much less than the kinetic energy scale, \textit{i.e.}, when $k_{B}T_{0} \ll \hbar^{2}\rho_{0}(0)^{2}/m$. Here, $R_{\text{TF}}=2N/\pi\rho_{0}(0)=\sqrt{2N} l_{ho}$ is the TF size of the cloud, with harmonic oscillator length $l_{ho}=\sqrt{\hbar/m\omega_{0}}$. Hence, Eq. (\ref{rho_{0}_TF}) together with Eq. (\ref{hydro_rho}) provide the low-temperature solution for the density profile of the TG gas within the hydrodynamic theory.

The momentum distribution of the TG gas is somewhat more involved to obtain. One first employs the local density approximation (LDA) and defines $\bar{n}(k;\rho,T)$ to be the equilibrium momentum distribution of a uniform TG gas of density $\rho$ at temperature $T$, normalised to $\rho$ (\textit{i.e.}, $\int dk~\bar{n}(k;\rho,T) = \rho$). In this approach, the momentum distribution of each uniform slice of gas can be added together to obtain the full momentum distribution for the trapped TG gas,
\begin{align}
	n(k,t) &= \int dx\, \bar{n}[k-mv(x,t)/\hbar; \rho(x,t), T(t)]. \label{total_n(k,t)_expression}
\end{align}
Next, one models the momentum distribution of a uniform TG gas using a Lorentzian $\bar{n}(k;\rho,T) = (2\rho l_{\phi}/\pi)/[1 + (2l_{\phi}k)^{2}]$ \cite{Bouchoule2016,Atas2017b}, where $l_{\phi}(x,t) = \hbar^{2}\rho(x,t)/mk_{B}T(t)$ is the phase coherence length of the gas. Although this model is only valid for small momenta and low temperatures ($|k| \ll 1/l_{\phi} \ll \rho$), it captures well the bulk of $\bar{n}$ and provides the dominant contribution to the bulk of $n(k,t)$. Using this Lorentzian approximation for $\bar{n}(k;\rho,T)$, the full momentum distribution of the TG gas is given by \cite{Bouchoule2016}
\begin{align}
	n(k,t) &= \frac{1}{\pi} \int dx\, \frac{2l_{\phi}(x,t)\rho(x,t)}{1 + 4[l_{\phi}(x,t)]^{2}[k-mv(x,t)/\hbar]^{2}}. \label{total_n(k,t)}
\end{align}

Defining the initial phase coherence length in the trap centre to be $l_{\phi}^{(0)} = \hbar^{2}\rho_{0}(0)/mk_{B}T(0)$ and applying the transformation $u=x/\lambda R_{\text{TF}}$ allows the momentum distribution to be written in dimensionless form as \cite{Bouchoule2016}
\begin{align}
	\frac{n(k,t)}{N l_{\phi}^{(0)}} &= \frac{4 \tilde{\lambda}}{\pi^{2}} \int_{-1}^{1} du\, \frac{(1-u^{2})}{1 + 4\tilde{\lambda}^{2}(1-u^{2})\left(\tilde{k}-\frac{2\pi}{\tilde{T}_{0}} \frac{\Omega}{\omega_{0}} \dot{\tilde{\lambda}} u\right)^{2}}, \label{n(k,t)_dimensionless}
\end{align}
where $\tilde{k}=kl_{\phi}^{(0)}$ and $\tilde{T}_{0}=T_{0}/T_{d}$, with $T_{d}=\hbar^{2}\rho_{0}(0)^{2}/2mk_{B}$ being the initial temperature of quantum degeneracy of a uniform 1D gas at density $\rho_{0}(0)$. In addition, $\tilde{\lambda}(\tau) \equiv \lambda(\tau/\Omega)$, where $\tau = \Omega t$ is the dimensionless time so that $\dot{\tilde{\lambda}} = d\tilde{\lambda}/d\tau = \dot{\lambda}(t)/\Omega$ \cite{Atas2017b}.

For the purpose of performing numerical calculations, one can use the relations $l_{\phi}^{(0)} = \hbar^{2}\rho_{0}(0)/mk_{B}T(0) = l_{ho}\sqrt{2/N}/\pi \theta_{0}$ and $\tilde{T}_{0} = \pi^{2} \theta_{0}$ (where $\theta_{0}=k_{B}T_{0}/N\hbar\omega_{0}$ is the initial dimensionless temperature of the gas) to rewrite Eq. (\ref{n(k,t)_dimensionless}) as
\begin{align}
	\frac{n(k,t)}{l_{ho}} &= C \tilde{\lambda} \int_{-1}^{1} du\, \frac{(1-u^{2})}{1 + B\tilde{\lambda}^{2}(1-u^{2})\left(\bar{k} - \sqrt{2N} \frac{\Omega}{\omega_{0}} \dot{\tilde{\lambda}} u\right)^{2}}. \label{n(k,t)_computational}
\end{align}
Here $C = 4\sqrt{2N}/\pi^{3}\theta_{0}$, $B = 8/\pi^{2}\theta_{0}^{2}N$, and $\bar{k} = k l_{ho}$. Note that, in order to construct the hydrodynamic momentum distribution given by Eq. (\ref{n(k,t)_computational}), only $N$, $\theta_{0}$ and the ratio $\Omega/\omega_{0}$ need to be specified.

\section{Exact expression of the Floquet exponent}
\label{Floquet-exp}

In this Appendix, we show how to obtain the exact expression for the Floquet exponent of Mathieu's equation, following the same line of derivation as in Ref.~\cite{mclachlan1964theory}. First, by inserting the Fourier expanded form of the periodic function $f(\tau)$, Eq.~(\ref{Fourier_Expansion}), into Mathieu's equation (\ref{Mathieu_eq}), one can find that the expansion coefficients $c_{n}$ satisfy the following three-term recurrence relation:
\begin{equation}
\left[(2n+\nu)^{2}-a\right]c_{n}+q(c_{n+1}+c_{n-1})=0. 
\label{recurrence_Fourier}
\end{equation}
This system of linear equations can be written in a matrix form,
\begin{equation}
(M_{\nu}-a\mathbb{1})\mathbf{c}=0,
\label{c-eqs}
\end{equation}
where the elements of the tridiagonal matrix $M_{\nu}$ are given by 
\begin{equation}
(M_{\nu})_{nm}=(2n+\nu)^2\delta_{nm}+q(\delta_{n+1,m}+\delta_{n-1,m}),
\label{M_matrix}
\end{equation}
and the vector $\mathbf{c}=(\dots, c_{-n}, \dots, c_{-1},c_{0},c_{1},\dots, c_{n}, \dots)$ contains the Fourier coefficients. 

The system of equations (\ref{c-eqs}) has a non-trivial solution if 
\begin{equation}
\mathrm{det}(M_{\nu}-a\mathbb{1})=0.
\label{determinant_Eq}
\end{equation}
This condition actually constitutes an equation for the Floquet exponent $\nu$, and requires computation of the infinite determinant. In practice, however, the infinite determinant must be truncated, so in order to ensure the convergence of the determinant, let us divide (\ref{recurrence_Fourier}) by $(2n+\nu)^2-a$ and define
\begin{equation}
	\zeta_{n}(\nu)=q/[(2n+\nu)^2-a].  \label{zeta}
\end{equation}

The determinant which appears in Eq. (\ref{determinant_Eq}) then becomes 
\begin{equation}
	\mathrm{det}\Big(\delta_{nm}+(\delta_{n+1,m}+\delta_{n-1,m})\zeta_{n}(\nu)\Big)\equiv \Delta(\nu) , \label{determinant_Delta}
\end{equation}
and setting $\Delta(\nu)=0$ will provide non-trivial solutions for the Floquet exponent $\nu$. 

Even in this form, the equation for $\nu$ remains rather intricate and difficult to solve. However, a closed form expression for the Floquet exponent can be obtained by making the following observations. First, since the value of a determinant does not change when rows are interchanged, it follows that the function $\Delta(\nu)$ is even [\textit{i.e.} $\Delta(\nu)=\Delta(-\nu)$]. Next, the transformation $\nu\rightarrow \nu\pm 2$ leaves the determinant unchanged since rows of the matrix are just shifted by $\pm 1$ in that case.
Hence, the function $\Delta(\nu)$ is periodic in $\nu$ with period $2$. The only singularities of $\Delta$ are simple poles located at $\nu_{n}=\pm \sqrt{a}-2n$, with $n\in \mathbb{Z}$. The function 
\begin{equation}
	\xi(\nu)=\frac{1}{\cos(\nu \pi)-\cos{(\sqrt{a}\pi)}}
\end{equation}
presents all the above mentioned properties, and by Liouville's theorem there exists a constant $\beta$ such that $\Delta (\nu)-\beta\xi(\nu)$ is also a constant \cite{mclachlan1964theory,arscott2014periodic}. A closed form expression for $\nu$ can now be obtained by determining this constant, as well as  $\beta$ by extension. These two constants will provide an equation for $\Delta(\nu)$ which can be set equal to zero and rearranged to obtain an expression for $\nu$.

Since $\Delta (\nu)-\beta\xi(\nu)$ is a constant, it can be evaluated at any $\nu$. Taking the limit $i\nu \rightarrow\infty$, we see that $\Delta(i\nu\rightarrow \infty)=1$, as all the off-diagonal elements $\zeta_{n}$ of the matrix in (\ref{determinant_Delta}) tend to zero in this limit, whereas the diagonal elements are equal to $1$, and $\xi(i\nu\rightarrow \infty)=0$. Consequently,
\begin{align}
	\Delta (\nu)-\beta\xi(\nu) = 1 \label{Liouville_equal_const}
\end{align}
and hence the constant $\beta$ is given by $\beta=(\Delta(\nu)-1)/\xi(\nu)$.

In particular, $\beta$ can be evaluated at $\nu=0$, provided that $a\neq 4j^{2}$ ($j\in \mathbb{Z}$) as to avoid the matrix elements $\zeta_{n}(0)$ becoming infinite for $n=j$, which results in $\beta = [1-\cos(\sqrt{a}\pi)][\Delta(0)-1]$. Substituting this into Eq. (\ref{Liouville_equal_const}) gives an explicit expression for the determinant,
\begin{equation}
	\Delta(\nu)=\frac{\Delta(0)\sin^{2}(\sqrt{a}\pi/2)-\sin^{2}(\nu \pi/2)}{\sin^{2}(\sqrt{a}\pi/2)-\sin^{2}(\nu \pi/2)}, \quad a\neq 4j^{2}. \label{Delta_nu_explicit}
\end{equation}
In this expression, $\Delta(0)$ is the determinant (\ref{determinant_Delta}) with the off-diagonal elements given by $\zeta_{n}(0)=q/(4n^2-a)$. Setting $\Delta(\nu) = 0$ (to obtain a non-trivial solution for the Floquet exponent) and rearranging for $\nu$ results in the following closed expression for the Floquet exponent:
\begin{equation}
	\nu =\frac{2}{\pi}\arcsin \left(\sqrt{\Delta(0)\sin^{2}(\sqrt{a}\pi/2)}\right), \quad a\neq 4j^{2}.
	\label{Floquet_Exponent_exact_a}
\end{equation}

Similarly, by evaluating $\beta$ at $\nu = 1$, it can be shown that, when $a=4j^{2}$,
\begin{equation}
	\nu =\frac{1}{\pi}\arccos \left(2\Delta(1)-1\right), \quad a= 4j^{2}.
	\label{Floquet_Exponent_exact_b}
\end{equation}
Here, $\Delta(1)$ is the determinant (\ref{determinant_Delta}) with the off-diagonal elements given by $\zeta_{n}(1)=q/((2n+1)^2-a)$.
In practice, one fixes the point in the parameter space $(a,q)$ and then computes the Floquet exponent $\nu$ using Eq.~(\ref{Floquet_Exponent_exact_a}) or (\ref{Floquet_Exponent_exact_b}), which are the same equations as Eqs.~(\ref{Floquet_Exponent_exact1}) and (\ref{Floquet_Exponent_exact2}) of the main text.

\section{Symmetry of the Fourier coefficients and form of the solutions in the different regions}
\label{Fourier-coeff}

Once the Floquet exponent is obtained, the Fourier coefficients can be computed from the eigenvector of the matrix $(M_{\nu}-a)$ which corresponds to the zero eigenvalue, where $M_{\nu}$ is given by Eq.~(\ref{M_matrix}).  From the exact expression of the Floquet exponent, (\ref{Floquet_Exponent_exact1}) or (\ref{Floquet_Exponent_exact2}), and the periodicity property of the determinant equation, Eq.~(\ref{determinant_Eq}), the value of $\nu$ can always be taken in the interval $[0,1]$ when it is real. In particular, the case $\nu=0,1$ corresponds to the integer values of the Floquet exponent.
The Floquet exponent can also be complex and it is easy to see that in that case the Fourier coefficients (\ref{recurrence_Fourier}) are also complex and lead to complex fundamental solutions (\ref{fundamental_solution}) and (\ref{second_fund_sol}). In practice, it is easier to work with real representations of the fundamental solutions. 
From the properties of the Floquet exponent outlined in the previous section, one can deduce symmetries for the Fourier coefficients and simplify the expression for the fundamental solutions in the different regions of the phase diagram (Fig.~\ref{Figure1}) in order to get a real representation. In what follows, we derive the exact expressions used in the numerical calculation of the two fundamental solutions in the different regimes of stability.

\subsection{Case when $\nu$ is complex ($\nu=1-i\eta$)}

When $(a,q)$ lies in between $b_{2n+1}(q)$ and $a_{2n+1}(q)$, the Floquet exponent has the form $\nu=1-i\eta$, which ultimately results in complex Fourier coefficients $c_{n}$. Consequently, the fundamental solution (\ref{fundamental_solution}) is not necessarily a real function. We can however use the properties of the Fourier coefficients to get a real solution. For that purpose, let us insert the form of the Floquet exponent $\nu=1-i\eta$ into the recurrence relation (\ref{recurrence_Fourier}). Setting $n \rightarrow -(n+1)$ and taking the complex conjugate of the resulting recurrence equation shows that the coefficients $c_{-n-1}^{\ast}$ satisfy the same recurrence relation as the coefficients $c_{n}$. This  means that they must be proportional to each other, \textit{i.e.}, $c_{-n-1}^{\ast}=(c_{-1}^{\ast}/c_{0})\, c_{n}$, where the proportionality constant is found by taking $n=0$, and that they must have the same complex amplitude, \textit{i.e.}, $|c_{-n-1}^{\ast}|^{2} = |c_{n}|^{2}$. Hence, the proportionality constant is a pure phase factor $c_{-1}^{\ast}/c_{0}=\exp(-2i\theta)$, and we finally have
\begin{equation}
	c_{n}=e^{2i\theta}c_{-n-1}^{\ast}.
\end{equation}

Using this result in the Fourier expansion (\ref{Fourier_Expansion}), we thus find that the fundamental solution (\ref{fundamental_solution}) takes the form
\begin{equation}
	z_{1}(\tau)=2e^{i\theta+\eta \tau}\sum_{n=0}^{\infty}\rho_{n}\cos((2n+1)\tau+\phi_{n}-\theta),
\end{equation}
where we have used the polar representation of the Fourier coefficients $c_{n}=\rho_{n}\exp(i\phi_{n})$ and where $2\theta=\phi_{0}+\phi_{-1}$. The solution is complex due to the presence of the phase factor $e^{i\theta}$. For real $\tau$ however, it is possible to absorb the phase factor into the definition of the fundamental solution $z_{1}\rightarrow z_{1}e^{-i\theta}$ (which remains a solution) in order to get a real solution
\begin{equation}
	z_{1}(\tau)=2e^{\eta \tau}\sum_{n=0}^{\infty}\rho_{n}\cos((2n+1)\tau+\phi_{n}-\theta). \label{First_Fundamental_solution_Unstable_Region1}
\end{equation}

The second fundamental solution in this region of the phase diagram is given by
\begin{equation}
	z_{2}(\tau)=2e^{-\eta \tau}\sum_{n=0}^{\infty}\rho_{n}\cos((2n+1)\tau-\phi_{n}+\theta).\label{Second_Fundamental_solution_Unstable_Region1}
\end{equation}

These two solutions are then combined according to (\ref{Cosine_elliptic}) and (\ref{Sine_elliptic}) to obtain the even and odd solutions, respectively, of Mathieu's equation.

\subsection{Case when $\nu$ is pure imaginary ($\nu=i\eta$)}

When $(a,q)$ lies between $b_{2n}(q)$ and $a_{2n}(q)$, the Floquet exponent is a pure imaginary number $\nu=i \eta$ with $\eta>0$ and therefore the Fourier coefficients are complex in this case as well. By inserting the Floquet exponent into the recurrence relation (\ref{recurrence_Fourier}) and setting $n\rightarrow -n$, along with the complex conjugate transformation, we deduce that the coefficients $c_{-n}^{\ast}$ are proportional to $c_{n}$. More precisely, we obtain $c_{n}=(c_{0}/c_{0}^{\ast})c_{-n}^{\ast}=\exp(2i\theta)c_{-n}^{\ast}$ with $\theta=\phi_{0}$. This property of the Fourier coefficients leads to the following form for the first fundamental solution:
\begin{equation}
	z_{1}(\tau)=e^{-\eta \tau}\left( \rho_{0}+2\sum_{n=1}^{\infty} \rho_{n}\cos(2n\tau+\phi_{n}-\theta)\right)
\end{equation}
where, as before, we have absorbed the extra phase coefficient $e^{i\theta}$ into the definition of the solution.

The second solution is obtained by making the substitution $\tau \rightarrow -\tau$,
\begin{equation}
	z_{2}(\tau)=e^{\eta \tau}\left( \rho_{0}+2\sum_{n=1}^{\infty} \rho_{n}\cos(2n\tau-\phi_{n}+\theta)\right).
\end{equation}

\subsection{Case when $\nu$ is real but not integer ($\nu \in \left]  0,1 \right[ $)}

When the pair $(a,q)$ lies in the stable region between $a_{n}(q)$ and $b_{n+1}(q)$, the Fourier coefficients are real and do not present any particular symmetry. The real valued even and odd solutions take the following form:
\begin{align}
	C(a,q,\tau)&=\sum_{n=-\infty}^{\infty}c_{n}\cos \left( (2n+\nu)\tau\right), \\
	S(a,q,\tau)&=\sum_{n=-\infty}^{\infty}c_{n}\sin \left( (2n+\nu)\tau\right),
\end{align}
with $0<\nu<1$. Notice that the second independent solution is obtained by replacing the cosine function by a sine function and that the normalization constant has been omitted in both representations.

\subsection{Case when $\nu$ is integer ($\nu=0,1$)}

As discussed above,  $\nu=0,1$ correspond to real integer values of the Floquet exponent and in this case only one solution can be periodic, in accordance with Ince's theorem \cite{mclachlan1964theory}. The second independent solution is constructed following a different scheme compared to the ones discussed previously.
When $(a,q)$ lies exactly on a characteristic line (dashed magenta line in Fig.~\ref{Figure1}), \textit{i.e.}, $a=a_{n}(q)$ or $a=b_{n}(q)$, the Fourier coefficients are all real. The fundamental $\pi$ or $2\pi$ periodic solution is even on $a_{n}$ and odd on $b_{n}$.

\subsubsection{On the characteristic lines of type $a_{n}(q)$}
When $a=a_{2n}(q)$, the $\pi$ periodic fundamental solution has the form
\begin{equation}
	C(a,q,\tau)=c_{0}+2\sum_{n=1}^{\infty}c_{n}\cos\left( 2 n \tau \right),
\end{equation}
which is proportional to the function commonly denoted in the literature as $ce_{2n}(\tau,q)$ \cite{mclachlan1964theory}. The second fundamental non-periodic solution is given by
\begin{align}
	\notag S(a,q,\tau)= & c_{0}\,\mathrm{Im}\left( J_{0}(\sqrt{q}e^{i \tau})Y_{0}(\sqrt{q}e^{-i \tau})\right) \\
	&+ 2\sum_{n=1}^{\infty} (-1)^{n} c_{n} \mathrm{Im}\left( J_{n}(\sqrt{q}e^{i \tau})Y_{n}(\sqrt{q}e^{-i \tau})\right),
\end{align}
where $J_{n}$ and $Y_{n}$ are the Bessel functions of first and second  kind, respectively. This function is commonly denoted as $fe_{2n}(\tau,q)$ in the literature \cite{mclachlan1964theory}. 

When $a=a_{2n+1}(q)$, the first fundamental solution has period $2\pi$ and can be written as
\begin{equation}
	C(a,q,\tau)=\sum_{n=0}^{\infty}c_{n}\cos\left( (2n+1)\tau\right),
\end{equation}
whereas the second solution takes the form
\begin{align}
	\notag  S(a,q,\tau)=\sum_{n=0}^{\infty}(-1)^{n}c_{n}\mathrm{Im}&\left[ J_{n}(\sqrt{q}e^{i \tau})Y_{n+1}(\sqrt{q}e^{-i \tau})\right. \\
	&\left. +J_{n+1}(\sqrt{q}e^{i \tau})Y_{n}(\sqrt{q}e^{-i \tau})\right].
\end{align}
In the literature, one can find the notation $ce_{2n+1}(\tau,q)$ and $fe_{2n+1}(\tau,q)$ for these two solutions, respectively.

\subsubsection{On the characteristic lines of type $b_{n}(q)$}
When $a=b_{2n+1}$, the is odd and one has 
\begin{equation}
	S(a,q,\tau)=\sum_{n=0}^{\infty}c_{n}\sin \left(  (2n+1)\tau\right),
\end{equation}
which is proportional to the commonly known function $se_{2n+1}(\tau,q)$ \cite{mclachlan1964theory}. The second solution has the form
\begin{align}
	\notag C(a,q,\tau)=\sum_{n=0}^{\infty} (-1)^{n}c_{n}\mathrm{Re}& \left[J_{n}(\sqrt{q}e^{i \tau})Y_{n+1}(\sqrt{q}e^{-i \tau})\right. \\
	& - \left. J_{n+1}(\sqrt{q}e^{i \tau})Y_{n}(\sqrt{q}e^{-i \tau})\right].
\end{align}
Finally, when $a=b_{2n+2}$ the solutions have the form
\begin{equation}
	S(a,q,\tau)=\sum_{n=0}^{\infty}c_{n+1}\sin\left( (2n+2)\tau\right),
\end{equation}
and 
\begin{align}
	\notag C(a,q,\tau)=&\sum_{n=0}^{\infty}  (-1)^{n}c_{n+1}\,\mathrm{Re}\left[J_{n}(\sqrt{q}e^{i \tau})Y_{n+2}(\sqrt{q}e^{-i \tau})\right. \\
	& - \left. J_{n+2}(\sqrt{q}e^{i \tau})Y_{n}(\sqrt{q}e^{-i \tau})\right].
\end{align}


\begin{thebibliography}{38}%
\makeatletter
\providecommand \@ifxundefined [1]{%
 \@ifx{#1\undefined}
}%
\providecommand \@ifnum [1]{%
 \ifnum #1\expandafter \@firstoftwo
 \else \expandafter \@secondoftwo
 \fi
}%
\providecommand \@ifx [1]{%
 \ifx #1\expandafter \@firstoftwo
 \else \expandafter \@secondoftwo
 \fi
}%
\providecommand \natexlab [1]{#1}%
\providecommand \enquote  [1]{``#1''}%
\providecommand \bibnamefont  [1]{#1}%
\providecommand \bibfnamefont [1]{#1}%
\providecommand \citenamefont [1]{#1}%
\providecommand \href@noop [0]{\@secondoftwo}%
\providecommand \href [0]{\begingroup \@sanitize@url \@href}%
\providecommand \@href[1]{\@@startlink{#1}\@@href}%
\providecommand \@@href[1]{\endgroup#1\@@endlink}%
\providecommand \@sanitize@url [0]{\catcode `\\12\catcode `\$12\catcode
  `\&12\catcode `\#12\catcode `\^12\catcode `\_12\catcode `\%12\relax}%
\providecommand \@@startlink[1]{}%
\providecommand \@@endlink[0]{}%
\providecommand \url  [0]{\begingroup\@sanitize@url \@url }%
\providecommand \@url [1]{\endgroup\@href {#1}{\urlprefix }}%
\providecommand \urlprefix  [0]{URL }%
\providecommand \Eprint [0]{\href }%
\providecommand \doibase [0]{http://dx.doi.org/}%
\providecommand \selectlanguage [0]{\@gobble}%
\providecommand \bibinfo  [0]{\@secondoftwo}%
\providecommand \bibfield  [0]{\@secondoftwo}%
\providecommand \translation [1]{[#1]}%
\providecommand \BibitemOpen [0]{}%
\providecommand \bibitemStop [0]{}%
\providecommand \bibitemNoStop [0]{.\EOS\space}%
\providecommand \EOS [0]{\spacefactor3000\relax}%
\providecommand \BibitemShut  [1]{\csname bibitem#1\endcsname}%
\let\auto@bib@innerbib\@empty
\bibitem [{\citenamefont {Girardeau}(1960)}]{Girardeau1960}%
  \BibitemOpen
  \bibfield  {author} {\bibinfo {author} {\bibfnamefont {M.}~\bibnamefont
  {Girardeau}},\ }\href {\doibase 10.1063/1.1703687} {\bibfield  {journal}
  {\bibinfo  {journal} {Journal of {M}athematical {P}hysics}\ }\textbf
  {\bibinfo {volume} {1}},\ \bibinfo {pages} {516} (\bibinfo {year}
  {1960})}\BibitemShut {NoStop}%
\bibitem [{\citenamefont {Paredes}\ \emph {et~al.}(2004)\citenamefont
  {Paredes}, \citenamefont {Widera}, \citenamefont {Murg}, \citenamefont
  {Mandel}, \citenamefont {F{\"o}lling}, \citenamefont {Cirac}, \citenamefont
  {Shlyapnikov}, \citenamefont {H{\"a}nsch},\ and\ \citenamefont
  {Bloch}}]{paredes2004tonks}%
  \BibitemOpen
  \bibfield  {author} {\bibinfo {author} {\bibfnamefont {B.}~\bibnamefont
  {Paredes}}, \bibinfo {author} {\bibfnamefont {A.}~\bibnamefont {Widera}},
  \bibinfo {author} {\bibfnamefont {V.}~\bibnamefont {Murg}}, \bibinfo {author}
  {\bibfnamefont {O.}~\bibnamefont {Mandel}}, \bibinfo {author} {\bibfnamefont
  {S.}~\bibnamefont {F{\"o}lling}}, \bibinfo {author} {\bibfnamefont
  {I.}~\bibnamefont {Cirac}}, \bibinfo {author} {\bibfnamefont {G.~V.}\
  \bibnamefont {Shlyapnikov}}, \bibinfo {author} {\bibfnamefont {T.~W.}\
  \bibnamefont {H{\"a}nsch}}, \ and\ \bibinfo {author} {\bibfnamefont
  {I.}~\bibnamefont {Bloch}},\ }\href@noop {} {\bibfield  {journal} {\bibinfo
  {journal} {Nature}\ }\textbf {\bibinfo {volume} {429}},\ \bibinfo {pages}
  {277} (\bibinfo {year} {2004})}\BibitemShut {NoStop}%
\bibitem [{\citenamefont {Kinoshita}\ \emph {et~al.}(2004)\citenamefont
  {Kinoshita}, \citenamefont {Wenger},\ and\ \citenamefont
  {Weiss}}]{Kinoshita2004}%
  \BibitemOpen
  \bibfield  {author} {\bibinfo {author} {\bibfnamefont {T.}~\bibnamefont
  {Kinoshita}}, \bibinfo {author} {\bibfnamefont {T.}~\bibnamefont {Wenger}}, \
  and\ \bibinfo {author} {\bibfnamefont {D.~S.}\ \bibnamefont {Weiss}},\ }\href
  {\doibase 10.1126/science.1100700} {\bibfield  {journal} {\bibinfo  {journal}
  {Science}\ }\textbf {\bibinfo {volume} {305}},\ \bibinfo {pages} {1125}
  (\bibinfo {year} {2004})}\BibitemShut {NoStop}%
\bibitem [{\citenamefont {Kinoshita}\ \emph {et~al.}(2006)\citenamefont
  {Kinoshita}, \citenamefont {Wenger},\ and\ \citenamefont
  {Weiss}}]{Kinoshita2006}%
  \BibitemOpen
  \bibfield  {author} {\bibinfo {author} {\bibfnamefont {T.}~\bibnamefont
  {Kinoshita}}, \bibinfo {author} {\bibfnamefont {T.}~\bibnamefont {Wenger}}, \
  and\ \bibinfo {author} {\bibfnamefont {D.~S.}\ \bibnamefont {Weiss}},\ }\href
  {\doibase 10.1038/nature04693} {\bibfield  {journal} {\bibinfo  {journal}
  {Nature}\ }\textbf {\bibinfo {volume} {440}},\ \bibinfo {pages} {900}
  (\bibinfo {year} {2006})}\BibitemShut {NoStop}%
\bibitem [{\citenamefont {Haller}\ \emph {et~al.}(2009)\citenamefont {Haller},
  \citenamefont {Gustavsson}, \citenamefont {Mark}, \citenamefont {Danzl},
  \citenamefont {Hart}, \citenamefont {Pupillo},\ and\ \citenamefont
  {N{\"a}gerl}}]{Naagerl2009}%
  \BibitemOpen
  \bibfield  {author} {\bibinfo {author} {\bibfnamefont {E.}~\bibnamefont
  {Haller}}, \bibinfo {author} {\bibfnamefont {M.}~\bibnamefont {Gustavsson}},
  \bibinfo {author} {\bibfnamefont {M.~J.}\ \bibnamefont {Mark}}, \bibinfo
  {author} {\bibfnamefont {J.~G.}\ \bibnamefont {Danzl}}, \bibinfo {author}
  {\bibfnamefont {R.}~\bibnamefont {Hart}}, \bibinfo {author} {\bibfnamefont
  {G.}~\bibnamefont {Pupillo}}, \ and\ \bibinfo {author} {\bibfnamefont
  {H.-C.}\ \bibnamefont {N{\"a}gerl}},\ }\href {\doibase
  10.1126/science.1175850} {\bibfield  {journal} {\bibinfo  {journal}
  {Science}\ }\textbf {\bibinfo {volume} {325}},\ \bibinfo {pages} {1224}
  (\bibinfo {year} {2009})}\BibitemShut {NoStop}%
\bibitem [{\citenamefont {Meinert}\ \emph {et~al.}(2015)\citenamefont
  {Meinert}, \citenamefont {Panfil}, \citenamefont {Mark}, \citenamefont
  {Lauber}, \citenamefont {Caux},\ and\ \citenamefont
  {N\"agerl}}]{Naagerl2015}%
  \BibitemOpen
  \bibfield  {author} {\bibinfo {author} {\bibfnamefont {F.}~\bibnamefont
  {Meinert}}, \bibinfo {author} {\bibfnamefont {M.}~\bibnamefont {Panfil}},
  \bibinfo {author} {\bibfnamefont {M.~J.}\ \bibnamefont {Mark}}, \bibinfo
  {author} {\bibfnamefont {K.}~\bibnamefont {Lauber}}, \bibinfo {author}
  {\bibfnamefont {J.-S.}\ \bibnamefont {Caux}}, \ and\ \bibinfo {author}
  {\bibfnamefont {H.-C.}\ \bibnamefont {N\"agerl}},\ }\href {\doibase
  10.1103/PhysRevLett.115.085301} {\bibfield  {journal} {\bibinfo  {journal}
  {Phys. Rev. Lett.}\ }\textbf {\bibinfo {volume} {115}},\ \bibinfo {pages}
  {085301} (\bibinfo {year} {2015})}\BibitemShut {NoStop}%
\bibitem [{\citenamefont {Bloch}\ \emph {et~al.}(2008)\citenamefont {Bloch},
  \citenamefont {Dalibard},\ and\ \citenamefont
  {Zwerger}}]{Bloch-Dalibard-Zwerger}%
  \BibitemOpen
  \bibfield  {author} {\bibinfo {author} {\bibfnamefont {I.}~\bibnamefont
  {Bloch}}, \bibinfo {author} {\bibfnamefont {J.}~\bibnamefont {Dalibard}}, \
  and\ \bibinfo {author} {\bibfnamefont {W.}~\bibnamefont {Zwerger}},\ }\href
  {\doibase 10.1103/RevModPhys.80.885} {\bibfield  {journal} {\bibinfo
  {journal} {Rev. Mod. Phys.}\ }\textbf {\bibinfo {volume} {80}},\ \bibinfo
  {pages} {885} (\bibinfo {year} {2008})}\BibitemShut {NoStop}%
\bibitem [{\citenamefont {Cazalilla}\ and\ \citenamefont
  {Rigol}(2010)}]{CazalillaRigolNJP2010}%
  \BibitemOpen
  \bibfield  {author} {\bibinfo {author} {\bibfnamefont {M.~A.}\ \bibnamefont
  {Cazalilla}}\ and\ \bibinfo {author} {\bibfnamefont {M.}~\bibnamefont
  {Rigol}},\ }\href {http://stacks.iop.org/1367-2630/12/i=5/a=055006}
  {\bibfield  {journal} {\bibinfo  {journal} {New Journal of Physics}\ }\textbf
  {\bibinfo {volume} {12}},\ \bibinfo {pages} {055006} (\bibinfo {year}
  {2010})}\BibitemShut {NoStop}%
\bibitem [{\citenamefont {Polkovnikov}\ \emph {et~al.}(2011)\citenamefont
  {Polkovnikov}, \citenamefont {Sengupta}, \citenamefont {Silva},\ and\
  \citenamefont {Vengalattore}}]{polkovnikov2011}%
  \BibitemOpen
  \bibfield  {author} {\bibinfo {author} {\bibfnamefont {A.}~\bibnamefont
  {Polkovnikov}}, \bibinfo {author} {\bibfnamefont {K.}~\bibnamefont
  {Sengupta}}, \bibinfo {author} {\bibfnamefont {A.}~\bibnamefont {Silva}}, \
  and\ \bibinfo {author} {\bibfnamefont {M.}~\bibnamefont {Vengalattore}},\
  }\href {\doibase 10.1103/RevModPhys.83.863} {\bibfield  {journal} {\bibinfo
  {journal} {Rev. Mod. Phys.}\ }\textbf {\bibinfo {volume} {83}},\ \bibinfo
  {pages} {863} (\bibinfo {year} {2011})}\BibitemShut {NoStop}%
\bibitem [{\citenamefont {Cazalilla}\ \emph {et~al.}(2011)\citenamefont
  {Cazalilla}, \citenamefont {Citro}, \citenamefont {Giamarchi}, \citenamefont
  {Orignac},\ and\ \citenamefont {Rigol}}]{Cazalilla2011}%
  \BibitemOpen
  \bibfield  {author} {\bibinfo {author} {\bibfnamefont {M.~A.}\ \bibnamefont
  {Cazalilla}}, \bibinfo {author} {\bibfnamefont {R.}~\bibnamefont {Citro}},
  \bibinfo {author} {\bibfnamefont {T.}~\bibnamefont {Giamarchi}}, \bibinfo
  {author} {\bibfnamefont {E.}~\bibnamefont {Orignac}}, \ and\ \bibinfo
  {author} {\bibfnamefont {M.}~\bibnamefont {Rigol}},\ }\href {\doibase
  10.1103/RevModPhys.83.1405} {\bibfield  {journal} {\bibinfo  {journal} {Rev.
  Mod. Phys.}\ }\textbf {\bibinfo {volume} {83}},\ \bibinfo {pages} {1405}
  (\bibinfo {year} {2011})}\BibitemShut {NoStop}%
\bibitem [{lam()}]{lamacraft2012}%
  \BibitemOpen
  \href@noop {} {}\bibinfo {howpublished} {A. Lamacraft and J. Moore, in
  \emph{Ultracold bosonic and fermionic gases}, Vol. 5 (Contemporary Concepts
  in Condensed Matter Science), Eds. K. Levin, A. L. Fetter, and D. M.
  Stamper-Kurn (Elsevier, The Netherlands, 2012).}\BibitemShut {Stop}%
\bibitem [{\citenamefont {Husimi}(1953)}]{Husimi1953}%
  \BibitemOpen
  \bibfield  {author} {\bibinfo {author} {\bibfnamefont {K.}~\bibnamefont
  {Husimi}},\ }\href {\doibase 10.1143/ptp/9.4.381} {\bibfield  {journal}
  {\bibinfo  {journal} {Progress of Theoretical Physics}\ }\textbf {\bibinfo
  {volume} {9}},\ \bibinfo {pages} {381} (\bibinfo {year} {1953})}\BibitemShut
  {NoStop}%
\bibitem [{\citenamefont {Breuer}\ and\ \citenamefont
  {Holthaus}(1989)}]{Breuer1989}%
  \BibitemOpen
  \bibfield  {author} {\bibinfo {author} {\bibfnamefont {H.~P.}\ \bibnamefont
  {Breuer}}\ and\ \bibinfo {author} {\bibfnamefont {M.}~\bibnamefont
  {Holthaus}},\ }\href {\doibase 10.1007/BF01436579} {\bibfield  {journal}
  {\bibinfo  {journal} {Zeitschrift f{\"u}r Physik D Atoms, Molecules and
  Clusters}\ }\textbf {\bibinfo {volume} {11}},\ \bibinfo {pages} {1} (\bibinfo
  {year} {1989})}\BibitemShut {NoStop}%
\bibitem [{\citenamefont {He}\ \emph {et~al.}(2014)\citenamefont {He},
  \citenamefont {Brown}, \citenamefont {Haas},\ and\ \citenamefont
  {Rigol}}]{He-Brown2014}%
  \BibitemOpen
  \bibfield  {author} {\bibinfo {author} {\bibfnamefont {K.}~\bibnamefont
  {He}}, \bibinfo {author} {\bibfnamefont {J.}~\bibnamefont {Brown}}, \bibinfo
  {author} {\bibfnamefont {S.}~\bibnamefont {Haas}}, \ and\ \bibinfo {author}
  {\bibfnamefont {M.}~\bibnamefont {Rigol}},\ }\href {\doibase
  10.1103/PhysRevA.89.033634} {\bibfield  {journal} {\bibinfo  {journal} {Phys.
  Rev. A}\ }\textbf {\bibinfo {volume} {89}},\ \bibinfo {pages} {033634}
  (\bibinfo {year} {2014})}\BibitemShut {NoStop}%
\bibitem [{\citenamefont {Brown}(1991)}]{Brown1991}%
  \BibitemOpen
  \bibfield  {author} {\bibinfo {author} {\bibfnamefont {L.~S.}\ \bibnamefont
  {Brown}},\ }\href {\doibase 10.1103/PhysRevLett.66.527} {\bibfield  {journal}
  {\bibinfo  {journal} {Phys. Rev. Lett.}\ }\textbf {\bibinfo {volume} {66}},\
  \bibinfo {pages} {527} (\bibinfo {year} {1991})}\BibitemShut {NoStop}%
\bibitem [{\citenamefont {Hänggi}\ and\ \citenamefont
  {Zerbe}(1993)}]{Hanggi1993}%
  \BibitemOpen
  \bibfield  {author} {\bibinfo {author} {\bibfnamefont {P.}~\bibnamefont
  {Hänggi}}\ and\ \bibinfo {author} {\bibfnamefont {C.}~\bibnamefont
  {Zerbe}},\ }\href {\doibase 10.1063/1.44641} {\bibfield  {journal} {\bibinfo
  {journal} {AIP Conference Proceedings}\ }\textbf {\bibinfo {volume} {285}},\
  \bibinfo {pages} {481} (\bibinfo {year} {1993})}\BibitemShut {NoStop}%
\bibitem [{\citenamefont {Kohler}\ \emph {et~al.}(1997)\citenamefont {Kohler},
  \citenamefont {Dittrich},\ and\ \citenamefont
  {H\"anggi}}]{Kohler-Hanggi1997}%
  \BibitemOpen
  \bibfield  {author} {\bibinfo {author} {\bibfnamefont {S.}~\bibnamefont
  {Kohler}}, \bibinfo {author} {\bibfnamefont {T.}~\bibnamefont {Dittrich}}, \
  and\ \bibinfo {author} {\bibfnamefont {P.}~\bibnamefont {H\"anggi}},\ }\href
  {\doibase 10.1103/PhysRevE.55.300} {\bibfield  {journal} {\bibinfo  {journal}
  {Phys. Rev. E}\ }\textbf {\bibinfo {volume} {55}},\ \bibinfo {pages} {300}
  (\bibinfo {year} {1997})}\BibitemShut {NoStop}%
\bibitem [{\citenamefont {Zerbe}\ \emph {et~al.}(1994)\citenamefont {Zerbe},
  \citenamefont {Jung},\ and\ \citenamefont {H\"anggi}}]{Zerbe-Hanggi1994}%
  \BibitemOpen
  \bibfield  {author} {\bibinfo {author} {\bibfnamefont {C.}~\bibnamefont
  {Zerbe}}, \bibinfo {author} {\bibfnamefont {P.}~\bibnamefont {Jung}}, \ and\
  \bibinfo {author} {\bibfnamefont {P.}~\bibnamefont {H\"anggi}},\ }\href
  {\doibase 10.1103/PhysRevE.49.3626} {\bibfield  {journal} {\bibinfo
  {journal} {Phys. Rev. E}\ }\textbf {\bibinfo {volume} {49}},\ \bibinfo
  {pages} {3626} (\bibinfo {year} {1994})}\BibitemShut {NoStop}%
\bibitem [{\citenamefont {Thorwart}\ \emph {et~al.}(2000)\citenamefont
  {Thorwart}, \citenamefont {Reimann},\ and\ \citenamefont
  {H\"anggi}}]{Thorwart-Hanggi2000}%
  \BibitemOpen
  \bibfield  {author} {\bibinfo {author} {\bibfnamefont {M.}~\bibnamefont
  {Thorwart}}, \bibinfo {author} {\bibfnamefont {P.}~\bibnamefont {Reimann}}, \
  and\ \bibinfo {author} {\bibfnamefont {P.}~\bibnamefont {H\"anggi}},\ }\href
  {\doibase 10.1103/PhysRevE.62.5808} {\bibfield  {journal} {\bibinfo
  {journal} {Phys. Rev. E}\ }\textbf {\bibinfo {volume} {62}},\ \bibinfo
  {pages} {5808} (\bibinfo {year} {2000})}\BibitemShut {NoStop}%
\bibitem [{\citenamefont {Quinn}\ and\ \citenamefont
  {Haque}(2014)}]{QuinnHaque2014}%
  \BibitemOpen
  \bibfield  {author} {\bibinfo {author} {\bibfnamefont {E.}~\bibnamefont
  {Quinn}}\ and\ \bibinfo {author} {\bibfnamefont {M.}~\bibnamefont {Haque}},\
  }\href {\doibase 10.1103/PhysRevA.90.053609} {\bibfield  {journal} {\bibinfo
  {journal} {Phys. Rev. A}\ }\textbf {\bibinfo {volume} {90}},\ \bibinfo
  {pages} {053609} (\bibinfo {year} {2014})}\BibitemShut {NoStop}%
\bibitem [{\citenamefont {Mathieu}(1868)}]{Mathieu1868}%
  \BibitemOpen
  \bibfield  {author} {\bibinfo {author} {\bibfnamefont {E.}~\bibnamefont
  {Mathieu}},\ }\href@noop {} {\bibfield  {journal} {\bibinfo  {journal}
  {Journal de Math\'{e}matiques Pures et Appliqu\'{e}es}\ }\textbf {\bibinfo
  {volume} {13}},\ \bibinfo {pages} {137} (\bibinfo {year} {1868})}\BibitemShut
  {NoStop}%
\bibitem [{\citenamefont {McLachlan}(1964)}]{mclachlan1964theory}%
  \BibitemOpen
  \bibfield  {author} {\bibinfo {author} {\bibfnamefont {N.}~\bibnamefont
  {McLachlan}},\ }\href {https://books.google.com.au/books?id=\_vEJjwEACAAJ}
  {\emph {\bibinfo {title} {Theory and Application of Mathieu Functions}}},\
  Dover Publications\ (\bibinfo  {publisher} {Dover},\ \bibinfo {year}
  {1964})\BibitemShut {NoStop}%
\bibitem [{\citenamefont {Atas}\ \emph
  {et~al.}(2017{\natexlab{a}})\citenamefont {Atas}, \citenamefont {Gangardt},
  \citenamefont {Bouchoule},\ and\ \citenamefont {Kheruntsyan}}]{Atas2017a}%
  \BibitemOpen
  \bibfield  {author} {\bibinfo {author} {\bibfnamefont {Y.~Y.}\ \bibnamefont
  {Atas}}, \bibinfo {author} {\bibfnamefont {D.~M.}\ \bibnamefont {Gangardt}},
  \bibinfo {author} {\bibfnamefont {I.}~\bibnamefont {Bouchoule}}, \ and\
  \bibinfo {author} {\bibfnamefont {K.~V.}\ \bibnamefont {Kheruntsyan}},\
  }\href {\doibase 10.1103/PhysRevA.95.043622} {\bibfield  {journal} {\bibinfo
  {journal} {Phys. Rev. A}\ }\textbf {\bibinfo {volume} {95}},\ \bibinfo
  {pages} {043622} (\bibinfo {year} {2017}{\natexlab{a}})}\BibitemShut
  {NoStop}%
\bibitem [{\citenamefont {Atas}\ \emph
  {et~al.}(2017{\natexlab{b}})\citenamefont {Atas}, \citenamefont {Bouchoule},
  \citenamefont {Gangardt},\ and\ \citenamefont {Kheruntsyan}}]{Atas2017b}%
  \BibitemOpen
  \bibfield  {author} {\bibinfo {author} {\bibfnamefont {Y.~Y.}\ \bibnamefont
  {Atas}}, \bibinfo {author} {\bibfnamefont {I.}~\bibnamefont {Bouchoule}},
  \bibinfo {author} {\bibfnamefont {D.~M.}\ \bibnamefont {Gangardt}}, \ and\
  \bibinfo {author} {\bibfnamefont {K.~V.}\ \bibnamefont {Kheruntsyan}},\
  }\href {\doibase 10.1103/PhysRevA.96.041605} {\bibfield  {journal} {\bibinfo
  {journal} {Phys. Rev. A}\ }\textbf {\bibinfo {volume} {96}},\ \bibinfo
  {pages} {041605} (\bibinfo {year} {2017}{\natexlab{b}})}\BibitemShut
  {NoStop}%
\bibitem [{\citenamefont {Fang}\ \emph {et~al.}(2014)\citenamefont {Fang},
  \citenamefont {Carleo}, \citenamefont {Johnson},\ and\ \citenamefont
  {Bouchoule}}]{Fang2014}%
  \BibitemOpen
  \bibfield  {author} {\bibinfo {author} {\bibfnamefont {B.}~\bibnamefont
  {Fang}}, \bibinfo {author} {\bibfnamefont {G.}~\bibnamefont {Carleo}},
  \bibinfo {author} {\bibfnamefont {A.}~\bibnamefont {Johnson}}, \ and\
  \bibinfo {author} {\bibfnamefont {I.}~\bibnamefont {Bouchoule}},\ }\href
  {\doibase 10.1103/PhysRevLett.113.035301} {\bibfield  {journal} {\bibinfo
  {journal} {Phys. Rev. Lett.}\ }\textbf {\bibinfo {volume} {113}},\ \bibinfo
  {pages} {035301} (\bibinfo {year} {2014})}\BibitemShut {NoStop}%
\bibitem [{\citenamefont {Bouchoule}\ \emph {et~al.}(2016)\citenamefont
  {Bouchoule}, \citenamefont {Szigeti}, \citenamefont {Davis},\ and\
  \citenamefont {Kheruntsyan}}]{Bouchoule2016}%
  \BibitemOpen
  \bibfield  {author} {\bibinfo {author} {\bibfnamefont {I.}~\bibnamefont
  {Bouchoule}}, \bibinfo {author} {\bibfnamefont {S.~S.}\ \bibnamefont
  {Szigeti}}, \bibinfo {author} {\bibfnamefont {M.~J.}\ \bibnamefont {Davis}},
  \ and\ \bibinfo {author} {\bibfnamefont {K.~V.}\ \bibnamefont
  {Kheruntsyan}},\ }\href {\doibase 10.1103/PhysRevA.94.051602} {\bibfield
  {journal} {\bibinfo  {journal} {Phys. Rev. A}\ }\textbf {\bibinfo {volume}
  {94}},\ \bibinfo {pages} {051602} (\bibinfo {year} {2016})}\BibitemShut
  {NoStop}%
\bibitem [{\citenamefont {Minguzzi}\ and\ \citenamefont
  {Gangardt}(2005)}]{GangardtMinguzziExact}%
  \BibitemOpen
  \bibfield  {author} {\bibinfo {author} {\bibfnamefont {A.}~\bibnamefont
  {Minguzzi}}\ and\ \bibinfo {author} {\bibfnamefont {D.~M.}\ \bibnamefont
  {Gangardt}},\ }\href {\doibase 10.1103/PhysRevLett.94.240404} {\bibfield
  {journal} {\bibinfo  {journal} {Phys. Rev. Lett.}\ }\textbf {\bibinfo
  {volume} {94}},\ \bibinfo {pages} {240404} (\bibinfo {year}
  {2005})}\BibitemShut {NoStop}%
\bibitem [{Erm()}]{Ermakov}%
  \BibitemOpen
  \href@noop {} {}\bibinfo {howpublished} {V. P. Ermakov, Univ. Izv. Kiev
  {\bf{20}}, 1 (1880).}\BibitemShut {Stop}%
\bibitem [{\citenamefont {Pinney}(1950)}]{Pinney}%
  \BibitemOpen
  \bibfield  {author} {\bibinfo {author} {\bibfnamefont {E.}~\bibnamefont
  {Pinney}},\ }\href {\doibase 10.1090/S0002-9939-1950-0037979-4} {\bibfield
  {journal} {\bibinfo  {journal} {Proc. Amer. Math. Soc.}\ }\textbf {\bibinfo
  {volume} {1}},\ \bibinfo {pages} {681} (\bibinfo {year} {1950})}\BibitemShut
  {NoStop}%
\bibitem [{\citenamefont {Lieb}\ and\ \citenamefont
  {Liniger}(1963)}]{Lieb-Liniger1963}%
  \BibitemOpen
  \bibfield  {author} {\bibinfo {author} {\bibfnamefont {E.~H.}\ \bibnamefont
  {Lieb}}\ and\ \bibinfo {author} {\bibfnamefont {W.}~\bibnamefont {Liniger}},\
  }\href {\doibase 10.1103/PhysRev.130.1605} {\bibfield  {journal} {\bibinfo
  {journal} {Phys. Rev.}\ }\textbf {\bibinfo {volume} {130}},\ \bibinfo {pages}
  {1605} (\bibinfo {year} {1963})}\BibitemShut {NoStop}%
\bibitem [{\citenamefont {Girardeau}\ and\ \citenamefont
  {Wright}(2000)}]{GirardeauWright2000}%
  \BibitemOpen
  \bibfield  {author} {\bibinfo {author} {\bibfnamefont {M.~D.}\ \bibnamefont
  {Girardeau}}\ and\ \bibinfo {author} {\bibfnamefont {E.~M.}\ \bibnamefont
  {Wright}},\ }\href {\doibase 10.1103/PhysRevLett.84.5691} {\bibfield
  {journal} {\bibinfo  {journal} {Phys. Rev. Lett.}\ }\textbf {\bibinfo
  {volume} {84}},\ \bibinfo {pages} {5691} (\bibinfo {year}
  {2000})}\BibitemShut {NoStop}%
\bibitem [{\citenamefont {Das}\ \emph {et~al.}(2002)\citenamefont {Das},
  \citenamefont {Girardeau},\ and\ \citenamefont {Wright}}]{Das-Girardeau2002}%
  \BibitemOpen
  \bibfield  {author} {\bibinfo {author} {\bibfnamefont {K.~K.}\ \bibnamefont
  {Das}}, \bibinfo {author} {\bibfnamefont {M.~D.}\ \bibnamefont {Girardeau}},
  \ and\ \bibinfo {author} {\bibfnamefont {E.~M.}\ \bibnamefont {Wright}},\
  }\href {\doibase 10.1103/PhysRevLett.89.170404} {\bibfield  {journal}
  {\bibinfo  {journal} {Phys. Rev. Lett.}\ }\textbf {\bibinfo {volume} {89}},\
  \bibinfo {pages} {170404} (\bibinfo {year} {2002})}\BibitemShut {NoStop}%
\bibitem [{\citenamefont {Popov}\ and\ \citenamefont
  {Perelomov}(1969)}]{popov1969parametric}%
  \BibitemOpen
  \bibfield  {author} {\bibinfo {author} {\bibfnamefont {V.}~\bibnamefont
  {Popov}}\ and\ \bibinfo {author} {\bibfnamefont {A.}~\bibnamefont
  {Perelomov}},\ }\href@noop {} {\bibfield  {journal} {\bibinfo  {journal}
  {Soviet Physics JETP}\ }\textbf {\bibinfo {volume} {30}},\ \bibinfo {pages}
  {910} (\bibinfo {year} {1969})}\BibitemShut {NoStop}%
\bibitem [{\citenamefont {Perelomov}\ and\ \citenamefont
  {Zel'dovich}(1998)}]{PerelomovBook}%
  \BibitemOpen
  \bibfield  {author} {\bibinfo {author} {\bibfnamefont {A.}~\bibnamefont
  {Perelomov}}\ and\ \bibinfo {author} {\bibfnamefont {Y.}~\bibnamefont
  {Zel'dovich}},\ }\href@noop {} {\emph {\bibinfo {title} {Quantum
  {M}echanics}}}\ (\bibinfo  {publisher} {World Scientific, Singapore},\
  \bibinfo {year} {1998})\BibitemShut {NoStop}%
\bibitem [{\citenamefont {Arscott}(2014)}]{arscott2014periodic}%
  \BibitemOpen
  \bibfield  {author} {\bibinfo {author} {\bibfnamefont {F.~M.}\ \bibnamefont
  {Arscott}},\ }\href@noop {} {\emph {\bibinfo {title} {Periodic differential
  equations: an introduction to Mathieu, Lam{\'e}, and allied functions}}},\
  Vol.~\bibinfo {volume} {66}\ (\bibinfo  {publisher} {Elsevier},\ \bibinfo
  {year} {2014})\BibitemShut {NoStop}%
\bibitem [{\citenamefont {Landau}\ and\ \citenamefont
  {Lifshitz}(1976)}]{landau1976mechanics}%
  \BibitemOpen
  \bibfield  {author} {\bibinfo {author} {\bibfnamefont {L.~D.}\ \bibnamefont
  {Landau}}\ and\ \bibinfo {author} {\bibfnamefont {E.~M.}\ \bibnamefont
  {Lifshitz}},\ }\href@noop {} {\emph {\bibinfo {title} {Mechanics}}}\
  (\bibinfo  {publisher} {Pergamon Press Oxford},\ \bibinfo {year}
  {1976})\BibitemShut {NoStop}%
\bibitem [{\citenamefont {Bell}(1957)}]{bell1957note}%
  \BibitemOpen
  \bibfield  {author} {\bibinfo {author} {\bibfnamefont {M.}~\bibnamefont
  {Bell}},\ }\href {\doibase 10.1017/S204061850003358X} {\bibfield  {journal}
  {\bibinfo  {journal} {Glasgow Mathematical Journal}\ }\textbf {\bibinfo
  {volume} {3}},\ \bibinfo {pages} {132} (\bibinfo {year} {1957})}\BibitemShut
  {NoStop}%
\bibitem [{\citenamefont {Kheruntsyan}\ \emph {et~al.}(2005)\citenamefont
  {Kheruntsyan}, \citenamefont {Gangardt}, \citenamefont {Drummond},\ and\
  \citenamefont {Shlyapnikov}}]{kheruntsyan2005}%
  \BibitemOpen
  \bibfield  {author} {\bibinfo {author} {\bibfnamefont {K.~V.}\ \bibnamefont
  {Kheruntsyan}}, \bibinfo {author} {\bibfnamefont {D.~M.}\ \bibnamefont
  {Gangardt}}, \bibinfo {author} {\bibfnamefont {P.~D.}\ \bibnamefont
  {Drummond}}, \ and\ \bibinfo {author} {\bibfnamefont {G.~V.}\ \bibnamefont
  {Shlyapnikov}},\ }\href {\doibase 10.1103/PhysRevA.71.053615} {\bibfield
  {journal} {\bibinfo  {journal} {Phys. Rev. A}\ }\textbf {\bibinfo {volume}
  {71}},\ \bibinfo {pages} {053615} (\bibinfo {year} {2005})}\BibitemShut
  {NoStop}%
\end{thebibliography}

%

\end{document}